\documentclass[12pt, pre,aps, showpacs,preprint]{revtex4}

\usepackage{amsmath}
\usepackage{amsfonts}
\usepackage{graphicx}
\usepackage{mathrsfs}
\usepackage{rotating}
\usepackage{setspace}
\usepackage{supertabular}
\usepackage{amsthm}
\usepackage{subfigure}
\usepackage{multirow}

\begin{document}
\title{Densest local packing diversity. II. Application to three dimensions}
\author{Adam B. Hopkins and Frank H. Stillinger}
\affiliation{Department of Chemistry, Princeton University, Princeton, New 
Jersey 08544}

\author{Salvatore Torquato}
\affiliation{Department of Chemistry, Department of Physics, Princeton Center
for Theoretical Science, 
Princeton Institute for the Science and Technology of Materials, Program in
Applied and Computational Mathematics, Princeton University, Princeton, New
Jersey 08544 \\ School of Natural Sciences, Institute for Advanced Study, 
Princeton, New Jersey 08544}

\begin{abstract}
The densest local packings of $N$ three-dimensional identical
nonoverlapping spheres within a radius $R_{min}(N)$ of a fixed central sphere of
the same size are obtained for selected values of $N$ up to $N=1054$. In the
predecessor to this paper [A.B. Hopkins, F.H. Stillinger and S. Torquato, Phys.
Rev. E {\bf 81} 041305 (2010)], we described our method for finding the putative
densest packings of $N$ spheres in $d$-dimensional Euclidean space ${\mathbb
R}^d$ and presented those packings in ${\mathbb R}^2$ for values of $N$ up to
$N=348$. We analyze the properties and characteristics of the densest local
packings in ${\mathbb R}^3$ and employ knowledge of the $R_{min}(N)$, using
methods applicable in any $d$, to construct both a realizability condition for
pair correlation functions of sphere packings and an upper bound on the {\it
maximal} density of {\it infinite} sphere packings. In ${\mathbb R}^3$, we find
wide variability in the densest local packings, including a multitude of packing
symmetries such as perfect tetrahedral and imperfect icosahedral symmetry. We
compare the densest local packings of $N$ spheres near a central sphere to
minimal-energy configurations of $N+1$ points interacting with short-range
repulsive and long-range attractive pair potentials, e.g., $12\!-\!6$
Lennard-Jones, and find that they are in general completely different, a result
that has possible implications for nucleation theory. We also compare the
densest local packings to finite subsets of stacking variants of the densest
infinite packings in ${\mathbb R}^3$ (the Barlow packings) and find that the
densest local packings are almost always most similar, as measured by a {\it
similarity metric}, to the subsets of Barlow packings with the smallest number
of coordination shells measured about a single central sphere, e.g., a subset of
the FCC Barlow packing. Additionally, we observe that the densest local packings
are dominated by the dense arrangement of spheres with centers at distance
$R_{min}(N)$. In particular, we find two ``maracas'' packings at $N=77$ and
$N=93$, each consisting of a few unjammed spheres free to rattle within a
``husk'' composed of the maximal number of spheres that can be packed with
centers at respective $R_{min}(N)$.

\end{abstract}

\maketitle

\section{Introduction}

A packing is defined as a set of nonoverlapping objects arranged in a space of
given dimension $d$. One packing problem in $d$-dimensional Euclidean space
${\mathbb R}^d$ that has not been generally addressed is that of finding the
maximally dense (optimal) packing(s) of $N$ nonoverlapping $d$-dimensional
spheres of unit diameter near (local to) an additional fixed central sphere such
that the greatest radius $R$ from any of the surrounding spheres' centers to the
center of the fixed sphere is minimized. This problem is called the densest
local packing (DLP) problem \cite{HST2010a}, and the minimized greatest radius
associated with number of spheres $N$ is denoted by $R_{min}(N)$. In various
limits, the DLP problem encompasses both the kissing number and (infinite)
sphere packing problems \cite{CSSPLG1998}. The former is a special case of the
DLP problem in that the kissing number $K_d$, or number of identical
$d$-dimensional nonoverlapping spheres that can simultaneously be in contact
with (kiss) a central sphere, is equal to the greatest $N$ for which $R_{min}(N)
= 1$, and the latter is equivalent to the DLP problem in the limit that $N
\rightarrow \infty$.

The DLP problem for $13$ spheres in ${\mathbb R}^3$ dates back to a debate
between Newton and Gregory in 1694. Newton believed that only $12$ identical
spheres could simultaneously contact a central same-size sphere, while Gregory
believed the correct number to be $13$. The first rigorous proof that Newton was
right came in 1953 \cite{SW1953a}, followed by a more concise proof in 1956
\cite{Leech1956a}. However, the question remains: how close can $13$ identical
spheres come to a central same-size sphere - how good was Gregory's guess? In
another paper \cite{HST2010a}, we showed that for any $d$, the DLP optimal
packings in ${\mathbb R}^d$ with $R_{min}(N) \leq \tau$, $\tau = (1+\sqrt{5})/2
\approx 1.618$ the golden ratio, include packings where all $N$ sphere centers
lie on a spherical surface of radius $R_{min}(N)$. The smallest radius spherical
surface onto which the centers of $13$ spheres of unit diameter can be placed is
strongly conjectured to be $R = R_{min}(13) = 1.045573\dots$, with the centers
arranged in a structure first documented in \cite{SW1951a}. It appears that
though Gregory was incorrect in conjecturing $K_3$ to be $13$, his guess wasn't
particularly far-off.

For each $N$ in the DLP problem in ${\mathbb R}^d$, there is a single optimal
$R_{min}(N)$, though generally for a given $N$ there can be multiple distinct
packings that achieve this radius. In the predecessor to this paper
\cite{HST2010b}, hereafter referred to as paper I, we studied the optimal
packings and corresponding $R_{min}(N)$ for the DLP problem in ${\mathbb R}^2$
for $N=1$ to $N=109$ and for $N$ corresponding to full shells of the triangular
lattice from $N=120$ to $N=348$. We also discussed the general concepts and
applications associated with DLP optimal packings for arbitrary $N$ and $d$.

In paper I, we reported that a majority of the DLP optimal packings in ${\mathbb
R}^2$ contain rattlers, or spheres (disks) that can be displaced in at least one
direction without increasing $R_{min}(N)$ or displacing any other sphere (disk)
in the packing. Further, we found that many optimal packings contain cavities at
their centers in which the central disk, were it not fixed, could move freely
(as a rattler, or object in a packing that can be displaced without displacing
any other objects or the packing boundary). The optimal packings in ${\mathbb
R}^2$ also exhibit a wide range of rotational symmetries, particularly for
smaller $N$, and packings of $N$ spheres from certain classes such as the curved
\cite{LG1997a} and wedge hexagonal packings were found to be optimal at various
$N$. We additionally observed that as $N$ grows large, disks in the {\it bulk}
(as opposed to on the surface) of DLP optimal packings are largely arranged as
subsets of the densest infinite packing in ${\mathbb R}^2$, i.e., with centers
on the sites of the triangular lattice, whereas disks farthest from the central
sphere (centers at distance $R_{min}(N)$) tend to form circular rings.

The DLP problem is related to problems of finding arrangements of $N+1$ points
${\bf r}^{N+1} \equiv {\bf r}_1, {\bf r}_2,\dots,{\bf r}_{N+1}$ that minimize
potential energy. Defining the DLP potential energy as the negative of the
density of $N+1$ spheres (including the central sphere) contained completely
within the {\it encompassing sphere}, a sphere of radius $R+1/2$ centered on the
central sphere, the DLP problem becomes a minimal-energy problem with the pair
potential between points (sphere centers) exhibiting features of long-range
attraction and infinite short-range repulsion. Comparisons between DLP optimal
packings and minimal-energy configurations for $N+1$ points with pair potentials
exhibiting similar features of long-range attraction and short-range repulsion
indicate that, though minimal energies for certain values of $N$ are similar,
optimal configurations of points (sphere centers) are in general completely
different. This finding could have implications for nucleation theory, as is
discussed in more detail in Secs. \ref{minEnergySection} and \ref{conclusions}.

The DLP problem is relevant to the realizability of functions that are
candidates to be the pair correlation function of a packing of identical
spheres. For a statistically homogeneous and isotropic packing, the pair
correlation function is denoted by $g_2(r)$; it is proportional to the
probability density of finding a separation $r$ between any two sphere centers
and normalized such that it takes the value of unity when no spatial
correlations between centers are present. Specifically, no function can be the
pair correlation function of a point process (where a packing of spheres of unit
diameter is a point processes with a minimum pair separation distance of unity)
unless it meets certain necessary, but generally not sufficient, conditions
known as realizability conditions \cite{Lenard1975a,TS2002a,KLS2007a}. Two of
these conditions that appear to be particularly strong for the realizability of
sphere packings \cite{TS2006a} are the nonnegativity of $g_2(r)$ and its
corresponding structure factor $S(k)$, where
\begin{equation}
S(k) = 1 + \rho \tilde{h}(k)
\label{structFactCond}
\end{equation}
with number density $\rho$ and $\tilde{h}(k)$ the $d$-dimensional Fourier
transform of the total correlation function $h(k) \equiv g_2(r)-1$.

Knowledge of the maximal number of sphere centers that may fit within radius $R$
from an additional fixed sphere's center, where that maximal number, denoted by
$Z_{max}(R)$, is equal to the greatest $N$ in the DLP problem for which
$R_{min}(N) \leq R$, may be employed to construct a third realizability
condition, called the $Z_{max}$ condition
\cite{HST2009a,HST2010a,CKT2009a,CE2003a}, on $g_2(r)$. This condition is
written
\begin{equation}
Z(R) \leq Z_{max}(R),
\label{ZRbound}
\end{equation}
where $Z(R)$ is defined for a statistically homogeneous packing as the expected
number of sphere centers within distance $R$ from an arbitrary sphere center.
The function $Z(R)$ can be related to the pair correlation function $g_2(r)$,
where for a packing of spheres with a pair correlation function $g_2({\bf r})$
that is direction-dependent, $g_2(r)$ is the directional average of $g_2({\bf
r})$, by
\begin{equation}
Z(R) = \rho s_1(1)\int_0^Rx^{d-1}g_2(x)dx.
\label{ZR}
\end{equation}
In Eq. (\ref{ZR}), $\rho$ is the constant number density of sphere centers and
$s_1(r)$ is the surface area of a sphere of radius $r$ in ${\mathbb R}^d$,
\begin{equation}
s_1(r) = \frac{2\pi^{d/2}r^{d-1}}{\Gamma(d/2)}.
\label{s1}
\end{equation}
As was discussed in previous papers \cite{HST2009a,HST2010a}, the $Z_{max}$
realizability condition has been shown to encode information not included in the
nonnegativity conditions on pair correlation functions and their corresponding
structure factors alone.

The maximal infinite-volume packing fraction $\phi_*^{\infty}$ of identical
nonoverlapping spheres in ${\mathbb R}^d$ can be bounded from above by employing
knowledge of the optimal $R_{min}(N)$ in the DLP problem, where a {\it packing
fraction} is the fraction of a given space covered by nonoverlapping objects. As
was discussed in paper I, an upper bound is constructed by measuring the packing
fraction $\phi(R+1/2)$ as the fraction of the volume of the encompassing sphere
covered by the $N+1$ spheres of unit diameter,
\begin{equation}
\phi(R+1/2) = \frac{N+1}{(2R+1)^d} = \rho\frac{\pi^{d/2}}{2^d\Gamma(1+d/2)}.
\label{packingFraction}
\end{equation}
In the first equality in (\ref{packingFraction}), $R$ is the greatest radius
from any of the $N$ surrounding spheres' centers to the center of the central
sphere, and in the second equality, the number density $\rho$ is the fraction of
a sphere of radius $\phi(R+1/2)$ covered by the $N+1$ spheres of unit diameter
and $\Gamma(x)$ is the standard gamma function.

For small numbers of spheres ($N \leq 1200$) in low dimensions ($d \leq 10$), an
algorithm combining a nonlinear programming method with a stochastic search of
configuration space can be employed on a personal computer to find putative
solutions to the DLP problem. The accuracy of the solutions found by such an
algorithm is bounded only by a machine's precision, and in general higher
accuracy only requires more computation time. Using such an algorithm, the
details of which are described in paper I, we find and present
putatively-optimal DLP packings and their corresponding $R_{min}(N)$ in
${\mathbb R}^3$ for $N = 1$ to $N=161$ (accuracy of at least $10^{-8}$ sphere
diameters), and for selected values of $N$ from $N=176$ to $N=1054$ (accuracy of
at least $10^{-6}$ sphere diameters). Images and coordinates for many of the
optimal packings that we have found are located on our website \cite{website}.
Of the $N\geq 176$ studied, some are randomly chosen and some correspond to the
numbers of contacting spheres of unit diameter near and equal to the number in
subsets of face-centered-cubic (FCC) and hexagonal-close-packed (HCP) packings
with a given number of full coordination shells.

In ${\mathbb R}^3$ as in ${\mathbb R}^2$, rigorous and repeated testing of the
algorithm indicates that it is robust in finding DLP optimal packings. The
identification of rotation and reflection symmetry, spatially precise to
$10^{-8}$ or better sphere diameters, in numerous DLP presumed-optimal packings
in ${\mathbb R}^3$ for $N\leq 114$ supports this conclusion, as does our finding
that the minimal $R$ found for $N=56$ to $N=58$ and for $N\geq 60$ are smaller
than the (previously) best-known minimal radii \cite{HugoWebsite} for the
less-restrictive problem of finding the minimal radius (larger) sphere into
which $N+1$ spheres of unit diameter may be packed. However, the algorithm is
dependent upon initial conditions, and as the number $N$ of spheres increases,
an increasing number of trials have been necessary before we have found a
packing and corresponding $R_{min}(N)$ that with a high degree of confidence we
consider optimal. For this reason, due to computing time constraints, though we
strongly conjecture that the vast majority of our packings are optimal for $N
\leq 161$, as many as $50\%$ or more of packings for $N \geq 176$ may not be
strictly optimal (though we hereafter refer to them as optimal). For these
packings, we will present evidence indicating that the smallest radius $R$ found
for each $N > 161$ is very near to the optimal $R_{min}(N)$, if not equal to it.

Over the range of $N$ studied, DLP optimal packings in ${\mathbb R}^3$ differ
substantially in terms of symmetry, contact networks, and spatial positioning
from the Barlow packings \cite{Barlow1883a}, which we recall are the maximally
dense infinite packings of identical nonoverlapping spheres in ${\mathbb R}^3$.
In general, we find that optimal packings are dominated by the dense arrangement
of spheres on their surface, i.e., those spheres with centers at precisely
distance $R_{min}(N)$, where the arrangement of the spheres in the bulk
(interior) is of secondary importance. Similarly in ${\mathbb R}^2$, DLP optimal
packings differ substantially from packings of contacting disks arranged on the
points of a triangular lattice, though in both dimensions at sufficiently large
$N$, the spheres in the bulk of each optimal packing begin to always resemble,
respectively, a subset of a corresponding maximally dense infinite packing in
${\mathbb R}^2$ or ${\mathbb R}^3$.

In ${\mathbb R}^2$, the approximate $N$ and corresponding $R_{min}(N)$ at which
this change occurs may be identified visually by perusing the DLP optimal
packings at various $N$. In ${\mathbb R}^3$ however, it can be difficult to
visually compare distinct packings. Consequently, we here introduce the concept
of a {\it similarity metric}, defined as a metric designed to quantify the
degree of similarity between one set of points and a reference set. As will be
discussed in Sec. \ref{SCandBarlow}, using such a metric allows quantitative
comparisons of the relative degree of similarity between the radial spatial
positions of spheres configured as DLP optimal packings and of spheres
configured as subsets of a maximally dense infinite packing.

Our key results and findings are summarized in the following list:
\begin{itemize}
\item A novel realizability condition on candidate pair correlation functions
$g_2(r)$ for sphere packings in ${\mathbb R}^3$ is constructed from knowledge of
the $R_{min}(N)$ (Sec. \ref{RealizabilityAndBounds}).
\item DLP optimal packings for almost every $N$ exhibit the phenomenon of
surface-maximization, i.e., the number of spheres {\it on the surface} (with
centers at precisely radius $R_{min}(N)$) is either the maximal or nearly the
maximal number that can be placed without overlap with centers on a spherical
surface of radius $R = R_{min}(N)$ (Sec. \ref{SCandBarlow1}). For two $N$
($N=77$ and $N=93$), the DLP optimal packings are termed ``maracas'' packings,
as packings at these $N$ have the maximal number of spheres on the surface,
while all spheres {\it not} on the surface are rattlers (Sec.
\ref{maracasSec}). 
\item DLP optimal packings for the $N$ studied are almost always {\it most}
similar to subsets of FCC Barlow packings, and for sufficiently large $N$, the
bulk (as opposed to the surface) of optimal packings appear to always be
structured similarly to a subset of a Barlow packing (Sec. \ref{SCandBarlow2}).
\item The set of $N$ for which there are DLP optimal packings that include
rattlers is unbounded, and the number of rattlers in a packing appears to grow
{\it at most} as quickly as the surface area of the packing (Sec.
\ref{SCandBarlow2}).
\item The $12\!-\!6$ Lennard-Jones (LJ) energy of DLP optimal packings for some
$N$ are within a few percent of the minimal LJ energy for $N+1$ points, but in
general the spatial configurations of sphere centers in DLP optimal packings are
completely different from the minimal-energy configurations of $N+1$ points
(Sec. \ref{minEnergySection}).
\item Many DLP optimal packings for $N\leq 114$ exhibit elements of perfect
rotational and/or reflection symmetry (Secs. \ref{denseSymm} and
\ref{otherSymm}).
\item Imperfect icosahedral symmetry is present in many DLP optimal packings,
but perfect icosahedral symmetry is {\it never} present (Sec.
\ref{icosPackings}).
\end{itemize}

\section{Realizability and bounds}
\label{RealizabilityAndBounds}
A function that is a candidate to be the pair correlation function $g_2(r)$ of a
point process, where a packing of spheres of unit diameter is a point process in
which the minimum pair distance is unity, must be nonnegative for all $r$ and
correspondingly have a structure factor $S(k)$ that is nonnegative for all $k$.
For a packing of spheres of unit diameter, such a candidate function must
additionally be identically zero on the interval $[0,1)$ to reflect the
nonoverlap condition between spheres. However, the two nonnegativity conditions
and the nonoverlap condition are only necessary, and generally not sufficient,
conditions for a function to be the pair correlation function of a sphere
packing. The $Z_{max}$ realizability condition further constrains candidate pair
correlation functions, eliminating a range of functions that obey the two
aforementioned nonnegativity conditions and are identically zero on $[0,1)$, and
yet violate the $Z_{max}$ condition \cite{HST2009a,HST2010a}.

\begin{figure}[ht]
\centering
\includegraphics[width = 3.2in,clip=true]{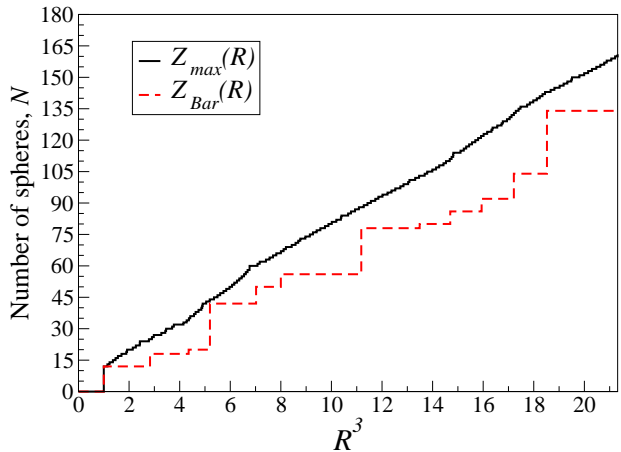}
\caption{(Color online) $Z_{max}(R)$ vs $R^3$, as determined by optimal and
putatively optimal solutions to the DLP problem for $N=1$ to $N=161$, and
$Z_{Bar}(R)$. The radius $R$ of the disk enclosing the centers of the $N$
identical (smaller) disks and same-size fixed disk is measured in units of the
diameter of the enclosed disks.} 
\label{ZmaxVsZBar}
\end{figure}

The function $Z_{max}(R)$ in ${\mathbb R}^3$ can be compared to $Z_{Bar}(R)$,
where $Z_{Bar}(R)$ is the largest number of spheres of unit diameter whose
centers can be placed within distance $R$ from a central sphere center when
packings are constrained to the space of Barlow packings. Both $Z_{max}(R)$ and
$Z_{Bar}(R)$ increase roughly linearly with $R^3$, as the volume of a sphere of
radius $R$ is proportional to $R^3$. The function $Z_{max}(R)$ is clearly always
greater than or equal to $Z_{Bar}(R)$, but we find strict inequality, i.e.,
$Z_{max}(R) > Z_{Bar}(R)$, for $N \geq 13$, as can be seen in Figure
\ref{ZmaxVsZBar}, a plot of $Z_{max}(R)$ and $Z_{Bar}(R)$ vs. $N$ for $N=1$ to
$N=161$.

\begin{figure}[ht]
\centering
\includegraphics[width = 3.2in,clip]{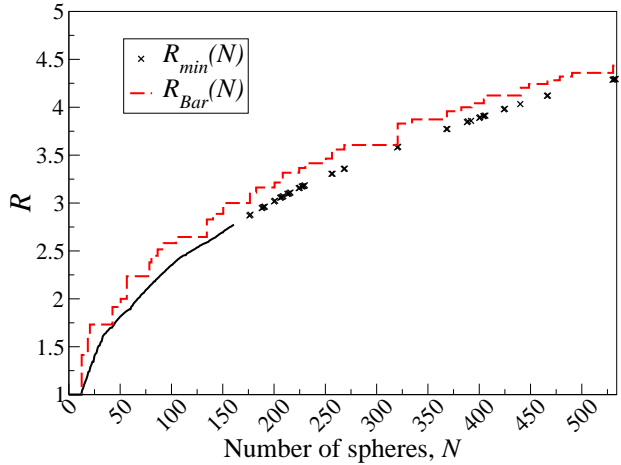}
\caption{(Color online) A plot of $R_{min}(N)$ vs. $N$ for $N=1$ to $N=161$ and
for selected values of $N \geq 176$ and $R_{Bar}(N)$ vs. $N$ for $N=1$ to
$N=533$.} 
\label{RminVsRBar}
\end{figure}

We also compare each $R_{min}(N)$ that we have found to each minimal radius
$R_{Bar}(N)$, where $R_{Bar}(N)$ is defined as the smallest $R$ for $N$ spheres
surrounding a fixed central sphere when packings are constrained to the space of
Barlow packings. Figure \ref{RminVsRBar} plots $R_{min}(N)$ and $R_{Bar}(N)$ vs.
$N$ for $N=1$ to $N=161$ and for the values of $N$ for which we used the
algorithm to seek optimal packings from $N=176$ to $N=533$. It is clear from the
figure that $R_{min}(N)$ rises roughly with $R_{Bar}(N)$, and that $R_{Bar}(N)$
is an upper bound for $R_{min}(N)$.

It is also true that $R_{Bar}(N)$ cannot grow too much larger than $R_{min}(N)$;
Appendix \ref{lowerBound} discusses methods of bounding $R_{min}(N)$ in
${\mathbb R}^3$ from below using the Barlow packings. We do not here explicitly
construct a lower bound for $R_{min}(N)$ that, along with the upper bound in
Fig. \ref{RminVsRBar}, could be used to specify a range of feasible values for
each optimal $R_{min}(N)$. However, we can comment on the accuracy of the
presumed-optimal $R_{min}(N)$ found by the algorithm for $N \geq 176$ by
comparing the differences between $R_{Bar}(N)$ and the putative $R_{min}(N)$ for
$N\geq 176$ and over the range $N \leq 161$ for which we have confidence in the
optimality of the packings. This comparison of $R_{Bar}(N) - R_{min}(N)$ over
the two ranges yields values for both the smallest and largest differences that
are very near to one another, as can be visually verified by close inspection of
Fig. \ref{RminVsRBar}.

In ${\mathbb R}^3$, knowledge of the Barlow packings allows us to bound
$R_{min}(N)$ and consequently $Z_{max}(R)$. However, in ${\mathbb R}^d$ with $d
> 3$ where the maximal infinite-volume packing fraction $\phi_*^{\infty}$ is
only known with analytical rigor for $d=8$ and $d=24$ \cite{CK2009a}, optimal
$R_{min}(N)$ can be employed to provide a rigorous upper bound on
$\phi_*^{\infty}$. As was shown in paper I, $\phi_*^{\infty}$ in ${\mathbb R}^d$
can be bounded from above with knowledge of any $R_{min}(N_*)$, where $N_* \in
{\mathbb N}$ is the set of all positive integers for which $N_*$ is the greatest
integer $N$ such that $R_{min}(N) = R_{min}(N_*)$. For example, in ${\mathbb
R}^3$, $R_{min}(1) = ... = R_{min}(12) = 1$, where $N_* = 12$ is the greatest
$N$ for which $R_{min}(N) = 1$ \cite{endnote4}.

The rigorous upper bound on $\phi_*^{\infty}$, proved in paper I, is
\begin{equation}
\phi_*^{\infty} \leq \hat{\phi}_*(N_*), \qquad N_* \in {\mathbb N},
\label{upperBound}
\end{equation}
where the maximal local packing fraction $\hat{\phi}_*(N)$ of a packing of $N$
nonoverlapping spheres of unit diameter around a same-size fixed central sphere
is defined as the ratio of the volumes of the $N+1$ spheres to the volume of a
sphere of radius $R_{min}(N)$, or

\begin{equation}
\hat{\phi}_*(N) = \frac{N+1}{(2R_{min}(N))^d}.
\label{maxLocalDensity}
\end{equation}

\begin{figure}[ht]
\centering
\includegraphics[width = 3.2in,clip]{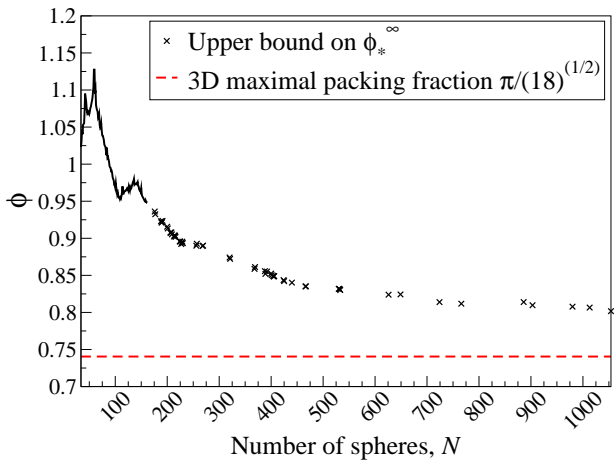}
\caption{(Color online) An upper bound on $\phi_*^{\infty}$ in ${\mathbb R}^3$
as calculated from the putative $R_{min}(N)$ for $N=34$ to $N=161$ and for
selected values of $N \geq 176$, versus the (known) maximal packing fraction
$\phi_*^{\infty}$ of an infinite packing of identical nonoverlapping spheres in
${\mathbb R}^3$, $\phi_*^{\infty} = \pi/\sqrt{18} \approx 0.7405$.}
\label{3DupperBoundPlot}
\end{figure}

Though $\phi_*^{\infty}$ is known rigorously in ${\mathbb R}^3$, it is
informative to calculate the bounds derived from the putative $R_{min}(N_*)$
over the range of $N$ tested. As $N$ increases, the bound becomes sharper
(becoming exact as $N \rightarrow \infty$), and calculations can give a sense of
how quickly the bound approaches the known value of $\phi_*^{\infty}$. Figure
\ref{3DupperBoundPlot} plots the upper bound calculated using relations
(\ref{upperBound}) and (\ref{maxLocalDensity}) for the putative $R_{min}(N_*)$
found in ${\mathbb R}^3$ versus the proved maximal infinite-volume packing
fraction, $\phi_*^{\infty} = \pi/\sqrt{18}$.

The upper bound in ${\mathbb R}^3$ (inequality (\ref{upperBound}), Fig.
\ref{3DupperBoundPlot}) with maximal local packing fraction $\hat{\phi}_*(N)$
calculated from Eq. (\ref{maxLocalDensity}) converges to $\phi_*^{\infty}$ more
slowly, as a function of $R_{min}(N)$, than does the upper bound in ${\mathbb
R}^2$ \cite{HST2010b}. This is intuitive. As the spheres in the bulk of each DLP
optimal packing in both ${\mathbb R}^2$ and ${\mathbb R}^3$ at sufficiently
large $N$ appear to be packed as the infinite densest packing in their
respective space ${\mathbb R}^d$ (this observation is discussed in more detail
in Sec. \ref{SCandBarlow2}), it would be logical to predict that
$\hat{\phi}_*(N)$ in the bound (\ref{upperBound}) converges to $\phi_*^{\infty}$
in dimension $d$ roughly as the ratio of sphere surface area to volume, $d/R$.
Up to a small corrective factor that is also dependent on dimension, this
appears to be the case, as the following simple model demonstrates.

The maximal local packing fraction $\hat{\phi}_*(N)$ (\ref{maxLocalDensity})
includes in its numerator the total volume of spheres of unit diameter with
centers at distance $R_{min}(N)$, even though the (larger) sphere of radius
$R_{min}(N)$ whose volume is employed in the denominator does not enclose
roughly half of these (smaller) spheres' volume. Approximating the spherical
surface of radius $R_{min}(N)$ as a plane, which is a good approximation when
the ratio of radii $1/2R_{min}(N)$ is small, a simple model for the volume of
the spheres not enclosed by the sphere of radius $R_{min}(N)$ can be built. In
this model under the aforementioned approximation, it follows that the
$d$-dimensional spheres of unit diameter in a DLP optimal packing in ${\mathbb
R}^d$ with centers on the surface at radius $R_{min}(N)$ should be packed
roughly as densely as possible, i.e., with centers arranged as the centers of
$(d\!-\!1)$-dimensional spheres in the maximally dense packing in ${\mathbb
R}^{d-1}$.

Making these two approximations, which become exact in the limit
$N\rightarrow\infty$, the fraction of the volume of the spheres of unit diameter
{\it not} enclosed by the sphere of radius $R_{min}(N)$ to the volume of the
sphere of radius $R_{min}(N)$ is $2/\sqrt{3} \approx 1.155$ times as large in
${\mathbb R}^3$ as in ${\mathbb R}^2$. At any $R_{min}(N)$ for large enough $N$
then, about $15.5\%$ more volume is ``added back in'' to the numerator of
$\hat{\phi}_*(N)$ in ${\mathbb R}^3$ than in ${\mathbb R}^2$, meaning that the
convergence fraction,
\begin{equation}
\phi_c(N) \equiv \phi_*^{\infty}/\hat{\phi}_*(N),
\label{phiC}
\end{equation}
at given $R_{min}(N)$ in ${\mathbb R}^3$ should be comparable to the same
fraction $\phi_c(N)$ at $(\sqrt{3}/2)R_{min}(N)$ in ${\mathbb R}^2$. This is
indeed the case, as is shown in Table \ref{convergeCompare} for two ranges of
$N$.

\begin{table}
\centering
\caption{Comparison of $\phi_c(N)$ (\ref{phiC}) in ${\mathbb R}^2$ and ${\mathbb
R}^3$. The values of $\phi_c(N)$ are compared for $R_{min}(N)$ in ${\mathbb
R}^2$ near to $(2/\sqrt{3})R_{min}(N)$ in ${\mathbb R}^3$.}
\begin{tabular}{c c c c c}
\hline
\hline
\,\,\,Space & $N$ & $R_{min}(N)$ & $(\sqrt{3}/2)R_{min}(N)$ & $\phi_c(N)$ \\
\hline
\,\,\,${\mathbb R}^2$ & $54\!-\!60$ & $3.605551\!-\!3.830649$ & &
$0.8733\!-\!0.9094$ \\
\,\,\,${\mathbb R}^3$ & $530\!-\!533$ & $4.286296\!-\!4.294254$ &
$3.712041\!-\!3.718933$ & $0.8785\!-\!0.8798$ \\
\hline
\,\,\,${\mathbb R}^2$ & $84\!-\!88$ & $4.581556\!-\!4.752754$ & &
$0.9065\!-\!0.9312$ \\
\,\,\,${\mathbb R}^3$ & \,\,\,$980$,$1013$,$1054$\,\,\, &
\,\,\,$5.334506\!-\!5.479129$\,\,\, & \,\,\,$4.619818\!-\!4.745065$\,\,\, &
$0.9167\!-\!0.9236$ \\
\hline
\end{tabular}
\label{convergeCompare}
\end{table}

The implications of this relationship are encouraging. Under the two
approximations, if the relationship between $\phi_c(N)$ in ${\mathbb R}^3$ and
${\mathbb R}^2$ holds between ${\mathbb R}^d$ and ${\mathbb R}^{d-1}$ for
arbitrary $d$, then a good estimate for $\phi_*^{\infty}$ in ${\mathbb R}^d$ can
be made. Such an estimate requires knowledge of $\phi_*^{\infty}$ in ${\mathbb
R}^{d-1}$ and ${\mathbb R}^{d-2}$ and at least one value of $R_{min}(N)$ for
sufficiently large $N$ in ${\mathbb R}^d$. In general, the larger the value of
$N$, the more precise the estimate.

\section{Comparing DLP optimal packings to optimal spherical codes and Barlow
packings}
\label{SCandBarlow}
In the first part of the following section, we compare DLP optimal packings over
the $N$ studied to the densest packings of $N$ nonoverlapping spheres of unit
diameter with centers on a spherical surface of radius $R$, where a
configuration of $N$ sphere centers with minimal $R = R^S_{min}$ is sometimes
called an optimal spherical code. In the second part of the section, we compare
DLP optimal packings to subsets of the densest infinite-volume packings of
identical nonoverlapping spheres in ${\mathbb R}^3$, the Barlow packings. 

\subsection{Spherical codes and DLP optimal packings}
\label{SCandBarlow1}
The spatial configurations of spheres in DLP optimal packings can be said to be
influenced by several empirical rules, but over the range of $N$ studied, they
are dominated by only one. The dominant rule is maximization of the number of
spheres on the {\it surface of the packing}, i.e., the spheres with centers at
precisely distance $R_{min}(N)$ from the center of the central sphere. In the
vast majority of DLP optimal packings over the range of $N$ studied, this number
$N_{out}$ is either the largest or nearly the largest number of spheres whose
centers can be placed on a spherical surface of radius $R_{min}(N)$. As
$N_{out}$ can take on different values in the general case where there are
multiple DLP optimal packings for the same $N$, we define $N_{out}(N)$ as the
maximal number of spheres, from the set of all DLP optimal packings for $N$
spheres, with centers at distance $R_{min}(N)$ from the center of the central
sphere. Similarly, we define $N^{Bar}_{out}(N)$ as the maximal number of spheres
with centers in the outermost coordination shell at distance $R_{Bar}(N)$ from a
central sphere surrounded by a subset of $N$ spheres chosen from a Barlow
packing. 

The number $N_{out}(N)$ for $R_{min}(N)$ is bounded from above by the number
$Z^S_{max}(R_{min}(N))$, where the maximal number of spheres of unit diameter
that can be packed with centers on a spherical surface of radius $R$ is termed
$Z^S_{max}(R)$. Related to $Z^S_{max}(R)$ is the radius $R^S_{min}(N)$, which we
recall is the radius of the smallest spherical surface onto which the centers of
$N$ nonoverlapping spheres of unit diameter can be packed. The problem of
finding $Z^S_{max}(R)$ at a given $R$ is a reformulation of the Tammes
\cite{Tammes1930a} problem of finding the maximal smallest separation between
pairs of points for $N$ points on a sphere of radius unity. The problem of
finding $R^S_{min}(N)$ is sometimes called the optimal spherical code problem
and has received considerable attention (see, for example, \cite{CSSPLG1998});
for $N\leq130$ and $d=3$, there are putative solutions to the optimal spherical
code problem that are strongly conjectured to be correct \cite{SloaneSCweb}.

\begin{figure}[ht]
\centering
\includegraphics[width = 3.2in,clip]{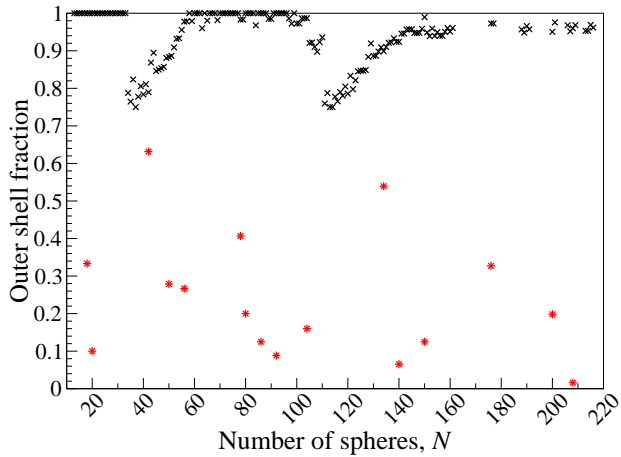}
\caption{(Color online) The greatest fraction, number of spheres with centers at
distance $R$ over the maximal number that can be packed with centers on a
spherical surface of radius $R$, as compared for all DLP optimal packings at a
given $N$ for the $N$ studied from $N=13$ to $N=216$ and for subsets of $N+1$
spheres in any Barlow packing with full coordination shells and the outermost
shell at distance $R_{Bar}(N)$. In the figure, a black ``X'' represents the
comparison for DLP optimal packings and a red ``*'' the comparison for subsets
of Barlow packings.}
\label{outerShellFrac}
\end{figure}

The quantity $N_{out}(N)/Z^S_{max}(R_{min}(N))$ is a measure of the degree to
which the surface of a DLP optimal packing is ``saturated'', where a {\it
saturated surface} of spheres in ${\mathbb R}^d$ is defined as any packing of
the maximal number $Z^S_{max}(R)$ of identical nonoverlapping spheres that can
be placed with centers at radius $R$. Using values for $Z^S_{max}(R)$ in
${\mathbb R}^3$ determined from \cite{SloaneSCweb} and the DLP optimal packings
found by the algorithm, Fig. \ref{outerShellFrac} compares
$N_{out}(N)/Z^S_{max}(R_{min}(N))$ and $N_{out}^{Bar}(N)/Z^S_{max}(R_{min}(N))$,
respectively, for the $N$ studied from $N=13$ to $N=216$ \cite{endnote2} and for
$N$ corresponding to full outermost coordination shells of Barlow packings with
outermost shells at radii $R_{Bar}(N)$.

For all of the $N$ studied, $N_{out}(N)/Z^S_{max}(R_{min}(N)) \geq 0.75$, and
$N_{out}(N) > N^{Bar}_{out}(N)$. Additionally, for $32\%$ of the $N$ in Fig.
\ref{outerShellFrac}, there is a DLP optimal packing with a saturated surface,
i.e., $N_{out}(N) = Z^S_{max}(R_{min}(N))$. These observations indicate that for
the $N$ studied, the densest local packings always include packings with a
maximal or near-maximal number of spheres on their surface. Nevertheless, there
are also two intervals depicted in Fig. \ref{outerShellFrac} where
$N_{out}(N)/Z^S_{max}(R_{min}(N))$ is relatively lower; these intervals coincide
with the existence at or near those $N$ of particularly dense DLP optimal
packings without saturated surfaces. In these intervals, the bulk of the packing
is of more relevance, and the benefits of maximizing the number of spheres on
the surface are relatively less advantageous to achieving a densest local
packing.

Though the surface-maximization rule is dominant in general, only for $N=77$
does $R^S_{min}(N_{out}(N)) = R_{min}(N)$. The $N=77$ DLP optimal packings are
therefore the only packings where the precise positions of the spheres {\it not}
on the surface are of no consequence. That is, of $N=77$ spheres, all $18$ not
on the surface are rattlers, which is also true for all $24$ spheres not on the
surface in the $N=93$ DLP optimal packings. Indeed as $N$ grows large, we have
found that the bulk (interior) of each DLP optimal packing begins to resemble a
finite subset of one of the Barlow packings, just as in ${\mathbb R}^2$ the bulk
of each optimal packing begins to resemble a finite subset of spheres configured
with centers on the sites of the triangular lattice.

\subsection{Barlow and DLP optimal packings}
\label{SCandBarlow2}
As discussed in paper I for large $N$ in ${\mathbb R}^2$, DLP optimal packings
can be divided into three regions; the spheres in the bulk that resemble the
triangular lattice, the spheres farthest from the center that tend to be
configured in circular rings, and those in between the bulk and the surface,
which form a sort of ``grain boundary.'' This also appears to be the case in
${\mathbb R}^3$, with the spheres in the bulk of the DLP optimal packings for
$N=766$, $903$, $980$, $1013$ and $1054$ closely resembling a subset of the FCC
Barlow packing ($N=903$, $980$, $1013$, $1054$) or resembling a packing that
near the central sphere is similar to a subset of the HCP Barlow packing
($N=766$).

It is of note that the spheres in the bulk of the aforementioned packings are
not all placed in {\it precisely} the same positions as spheres in a subset of a
Barlow packing, but instead are within a few percent of the Barlow packing
spheres' angular and radial positions as described in spherical coordinates. It
is not clear if this continues to be the case for all $N > 1054$, or if for some
very large $N$ the bulk of DLP optimal packings are precisely spatially
equivalent to subsets of Barlow packings.

The division of DLP optimal packings at sufficiently large $N$ into three
regions is important from the perspective of counting rattlers. For the five $N$
studied where the bulk of the optimal packings closely resemble subsets of
Barlow packings, rattlers are present only in the ``grain boundary'' and surface
regions. This suggests that for sufficiently large $N$, the rattlers in DLP
optimal packings are restricted to only these two regions. Consequently, as
these regions grow in volume more slowly than the volume of the packing, the
ratio of the number of rattlers to the total number of spheres in DLP optimal
packings must tend to zero at $N\rightarrow \infty$. Further, if the grain
boundary region for large enough $N$ does not increase in radial extent with
increasing $N$, then the number of rattlers can grow {\it at most} in proportion
to the surface area of optimal packings. This latter condition appears to be the
case for the $N$ studied in both ${\mathbb R}^2$ \cite{HST2010b} and ${\mathbb
R}^3$.

In ${\mathbb R}^2$, the extent to which the bulk of a DLP optimal packing
resembles a packing of contacting disks with centers on the sites of the
triangular lattice can be determined visually by perusing an image of the
packing. In ${\mathbb R}^3$ however, visual identification is more difficult. To
compare the relative extent to which the bulk of DLP optimal packings in
${\mathbb R}^3$ resemble subsets of the Barlow packings, we here introduce the
concept of a {\it similarity metric}, defined as a metric designed to quantify
the degree of similarity between one set of points and a reference set.

The subsets of $N+1$ spheres from a Barlow packing with the smallest distance
$R_{Bar}(N)$ from the center of the central sphere to any of the surrounding $N$
spheres are not always subsets of the FCC and HCP Barlow packings. This
observation suggests that we should compare the DLP optimal packings at various
$N$ to reference sets chosen from all Barlow packings, as opposed to only those
chosen from the FCC and HCP packings. Therefore, for each $N$, we define the set
of reference sets ${\mathbb B}_N$ as all subsets of $N+1$ spheres chosen from
any Barlow packing such that all $N+1$ spheres with centers less than or equal
to maximal sphere-distance $R$ from the center of a central sphere are included
in the set.

As rigid rotations of DLP optimal packings about the center of the central
sphere do not affect packing optimality, we employ a similarity metric that
compares only the radial positions of the spheres in an optimal packing to the
reference sets. To make this comparison for a DLP optimal packing and a packing
from the reference sets at a given $N$, which we recall are the Barlow packing
subsets ${\mathbb B}_N$, ${\mathbb R}^d$ is divided fully into a set
$\{\delta_i\}$ of nonoverlapping spherical shells centered on the center of the
reference set's central sphere. Each shell contains within it a number
$n^{ref}_i$ of points (sphere centers) from the reference set and a number $n_i$
of sphere centers from the set to be compared. The metric can be written,
\begin{equation}
{\mathcal S} = 1 - \frac{\sum_i|n_i-n^{ref}_i|}{2(N+1)},
\label{simMetric}
\end{equation}
where the sum runs over all shells containing at least one point from either
set.

For any Barlow packing subset of $N+1$ spheres within a given set ${\mathbb
B}_N$, the center of each sphere lies on a coordination shell, where we define
the coordination shells locally from the center of only the central sphere. For
example, the zeroeth shell is the origin and contains only the center of the
central sphere, and the first shell always contains the centers of $12$ spheres
at distance unity from the origin. To define the radial width of the shells
$\{\delta_i\}$ by which the $n_i$ and $n_i^{ref}$ are measured, an average of
the radial distances of consecutive coordination shells in the reference packing
is used. For example, for an FCC packing with coordination shells at
$r=0,1,\sqrt{2},\dots$, the zeroeth shell is the sphere of radius $(1/2)$
centered at the origin, and the first shell $\delta_1$ of radial width
$1/\sqrt{2}$ spans from minor radius $1/2$ to major radius $(1+\sqrt{2})/2$. The
final shell can be taken to have infinite width. More detailed information on
the reference sets and the choice of the similarity metric defined in Eq.
(\ref{simMetric}) can be found in Appendix \ref{finiteBarlowPackings}.

\begin{figure}[ht]
\centering
\subfigure[]{\includegraphics[width=3.2in,clip]{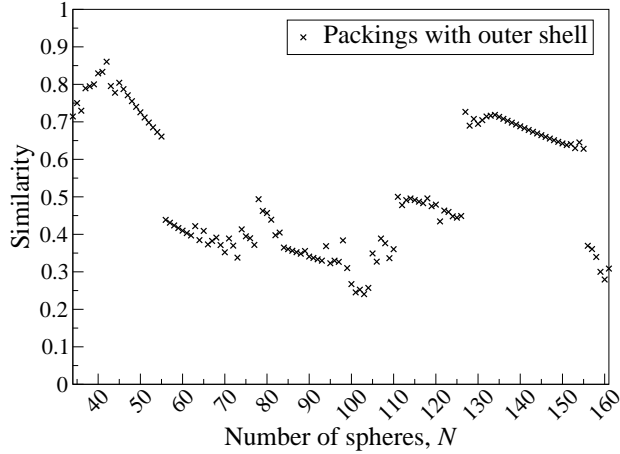}} \\
\subfigure[]{\includegraphics[width=3.2in,clip]{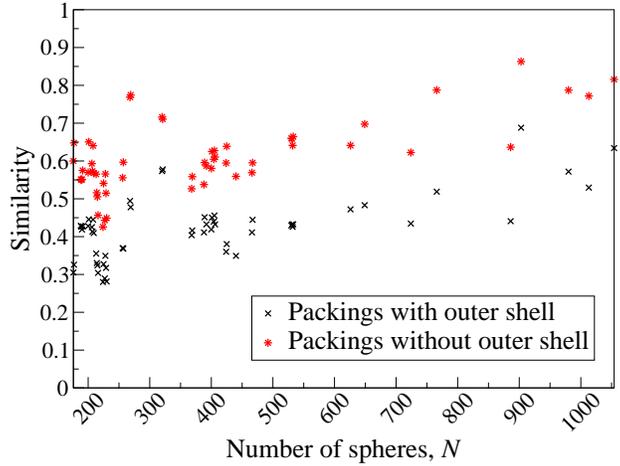}}
\caption{(Color online) Similarity metric from (\ref{simMetric}). An ``X''
represents the maximal similarity calculated for the comparisons of reference
sets ${\mathbb B}_N$ to a DLP optimal packing for $N$ spheres. A ``*''
represents the maximal similarity for the comparisons of reference sets
${\mathbb B}_M$ to a DLP optimal packing for $N$ spheres with the $N-M$ spheres
with centers at distance $R_{min}(N)$ removed.}
\label{maxSimMetric}
\end{figure}

Figure \ref{maxSimMetric} plots the greatest similarity metric value from the
distinct $\{\delta_i\}$ for all numbers $N$ of spheres studied, along with the
greatest similarity metric value for DLP optimal packings for $176 \leq N \leq
1054$ with all spheres with centers at distance $R_{min}(N)$ removed. As is
evident from the bottom half of Fig. \ref{maxSimMetric}, there is a gradual
upward trend in values of ${\mathcal S}$; this is primarily due to the bulk of
DLP optimal packings beginning to closely resemble subsets of a Barlow packings
for $N\geq 626$, $R_{min}(626) = 4.564905$. In particular, all but one or two of
the first $200$ spheres in the $N=766$, $903$, $980$, $1013$ and $1054$ optimal
packings are arranged in precisely the same first $10$ shells $\{\delta_1 \dots
\delta_{10}\}$ as are the spheres that compose one of the $N=200$ finite subsets
of Barlow packings ${\mathbb B}_{200}$.

However, only a few DLP optimal packings for $N\leq 533$, $R_{min}(533) =
4.294254$ bear close resemblance to any of those in the sets ${\mathbb B}_N$.
This result is due to the dominance of the empirical rule of
surface-maximization in influencing the spatial arrangements of spheres in DLP
optimal packings. Simply put, the surface-maximization phenomenon disrupts the
placement, sufficiently near to the surface, of spheres as subsets of Barlow
packings. The range of $R_{min}(N)$ in ${\mathbb R}^3$ over which this
disruption is prevalent throughout the entirety of optimal packings is
consistent with the same range of $R_{min}(N)$ for DLP optimal packings in
${\mathbb R}^2$, where we compare, as in Table \ref{convergeCompare}, values of
$(\sqrt{3}/2)R_{min}(N)$ in ${\mathbb R}^3$ to values of $R_{min}(N)$ in
${\mathbb R}^2$. In ${\mathbb R}^2$, signs of the bulk of a DLP optimal packing
resembling a triangular lattice packing of contacting disks for consecutive $N$
appear at the earliest around $N=76$, $R_{min}(76) = 4.417162\dots$, and do not
appear consistently until at least $N \geq 102$, with $R_{min}(102) =
5.166450\dots$ \cite{HST2010b}.

Included in the subsets of Barlow packings ${\mathbb B}_N$ are always packings
derived from the FCC and HCP lattices. The former of these is particularly
important, as we have found that the highest value of the similarity metric for
DLP optimal packings from among the distinct $\{\delta_i\}$ for $169$ out of
$184$, or $91.8\%$ of the $N \geq 34$ that we analyzed, is associated with the
FCC-derived packing and its variants indistinguishable to the similarity metric
\cite{endnote5}, even when $R_{Bar}(N)$ is found from a different packing. This
is important, as the FCC-derived packing and its variants indistinguishable to
the similarity metric are the packings that, for all $N$ in any set ${\mathbb
B}_N$, have the fewest coordination shells.

We have verified for the $N$ studied that the DLP optimal packings {\it also}
consist of shells (of small radial width) containing sphere centers clustered
around a relatively low number of radial distances from the center of the
central sphere. This is the reason that the highest similarity metric value for
all DLP optimal packings chosen from the reference sets in a given ${\mathbb
B}_N$ is almost always, for the $N$ studied, associated with the packings that
are subsets of an FCC packing: FCC and DLP optimal packings have small numbers
of shells.

This characteristic of DLP optimal packings is related to the phenomenon of
surface-maximization. The high density of spheres on the surface, due to
nonoverlap, requires that the radial separation between the surface spheres and
contacting spheres with centers at $r < R_{min}(N)$ be, in general, relatively
larger than if the surface contained fewer spheres. Coupled with
density-maximization, the rule of surface-maximization drives the spheres for $r
< R_{min}(N)$ to cluster around only a few distinct radial positions as well.
This observation is reflected in the fact that ${\mathcal S}$ for $48$ out of
$56$, or $85.7\%$ of packings excluding the spheres with centers at $R_{min}(N)$
for $N \geq 176$ are also most similar to the FCC-derived Barlow packing and its
variants indistinguishable to the similarity metric. However, the presence of
the $N=766$ packing, which exhibits a bulk that is very similar to a Barlow
packing that is not FCC, suggests that this trend may not continue as strongly
as $N$ grows larger than $N=1054$.

\begin{figure}[ht]
\centering
\includegraphics[width = 3.2in,clip]{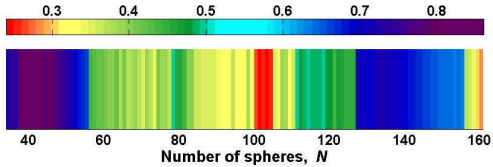}
\caption{(Color online) Similarity metric from (\ref{simMetric}), color-coded as
indicated in the key above the diagram, with violet/blue representing the
highest values of ${\mathcal S}$ and red/orange the lowest. The value of
${\mathcal S}$ displayed is calculated as the maximum for DLP optimal packings
for a given $N$ compared to the reference sets included in ${\mathbb B}_N$. Due
to the similarity between subsets of FCC packings and certain packings with
icosahedral symmetry, the ranges of $N$ with highest ${\mathcal S}$, $34 \leq N
\leq 55$ and $127 \leq N \leq 155$, are radially and angularly distributed most
similarly to a specific, dense packing of spheres with perfect icosahedral
symmetry.}
\label{colorBar}
\end{figure}

Despite that DLP optimal packings for the $N$ studied are generally more
radially similar to subsets of FCC than to subsets of other Barlow packings, for
$N\leq 533$ they are not angularly similar to any packings in ${\mathbb B}_N$.
For optimal packings with high values of ${\mathcal S}$, e.g., for $34 \leq N
\leq 55$ and $127 \leq N \leq 155$, the similarity is due entirely to the
arrangement of spheres in small numbers of shells, and the DLP optimal packings
are more similar to certain dense packings exhibiting icosahedral symmetry.
Figure \ref{colorBar}, a color-coded representation of the maximal ${\mathcal
S}$ from among the reference sets ${\mathbb B}_N$ for $34 \leq N \leq 161$,
represents well these regions of icosahedral symmetry. The dense, perfectly
icosahedrally symmetric packings to which these DLP optimal packings are similar
will be discussed in more detail in Sec. \ref{icosPackings}.

It is very interesting to note that an FCC arrangement of identical
nonoverlapping spheres, in the limit as infinite volume packing fraction
$\phi^{\infty}\rightarrow \phi_*^{\infty} = \pi/\sqrt{18}$, is both the Barlow
packing with highest symmetry (cubic) and lowest free energy \cite{MH1999a}. Our
results state that the densest local packings are most frequently those that are
most similar to the maximally dense infinite packing with highest symmetry and
lowest free energy even when other packings in ${\mathbb B}_N$ are more locally
dense than the FCC-derived packing, i.e., have smaller $R_{Bar}(N)$. As the
correspondence in similarity is essentially due to the arrangement of spheres in
DLP optimal packings in a relatively small number of shells, this suggests that
there is a connection between high symmetry, lowest free energy, and arrangement
in a small number of shells in the densest local packings, just as there is
between high symmetry, lowest free energy, and arrangement in a small number of
shells in the densest infinite packings.

\section{Minimal energy and DLP problems}
\label{minEnergySection}
There have been a number of investigations
\cite{HP1971a,Northby1987a,DW1995a,WD1997a,XJCS2004a,CambridgeCluster} into
finding arrangements of points that minimize the $12\!-\!6$ Lennard-Jones
potential, a potential possessing features of long-range attraction and strong
short range repulsion between pairs of points. The Lennard-Jones potential
energy for $N+1$ points can be written,
\begin{equation}
V_{LJ}({\bf r}^{N+1})=4\epsilon\!\!\!\!\!\!\!\sum_{1\leq i < j \leq
N+1}\left[\left(\!\frac{\sigma}{r_{ij}}\!\right)^{12} -
\left(\!\frac{\sigma}{r_{ij}}\!\right)^6\right],
\label{LJenergy}
\end{equation}
with $r_{ij} \equiv |{\bf r}_i-{\bf r}_j|$ and where the parameters $\epsilon$
and $\sigma$ can be set $\epsilon=\sigma=1$. Lennard-Jones minimal-energy
(optimal) configurations of $N+1$ points in ${\mathbb R}^3$ can be compared to
DLP optimal packings of $N$ spheres around a fixed central sphere of the same
size. This comparison is accomplished by scaling the DLP optimal packings such
that the minimal distance between sphere centers $D$ is optimized to minimize
the Lennard-Jones potential energy given by (\ref{LJenergy}) with $\epsilon =
\sigma = 1$.

\begin{figure}[ht]
\centering
\includegraphics[width = 3.2in,clip]{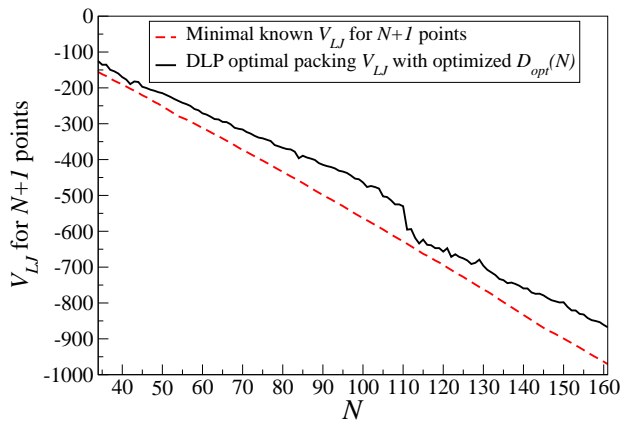}
\caption{(Color online) Plot of the minimum known \cite{CambridgeCluster}
$V_{LJ}({\bf r}^{N+1})$ (\ref{LJenergy}) for $N+1$ points vs the $V_{LJ}$ of DLP
optimal packings with spheres of optimized diameter $D_{opt}(N)$ for $N=34$ to
$N=161$.}
\label{DLPLJ}
\end{figure}

The optimal sphere diameters $D_{opt}(N)$ for $N=34$ to $N=161$ lie within a
tight range, between $D_{opt}(160) = 1.07953$ and $D_{opt}(44) = 1.09345$, and
they average about $1.08319$. These diameters may be compared to the
Lennard-Jones pair potential minimum, $D=2^{1/6} \approx 1.12246$. The
$D_{opt}(N)$ tend to decrease with increasing $N$, reflecting a balance obtained
as the packing is scaled between the increase in energy due to spheres in
contact at distance $D_{opt}(N) < 2^{1/6}$ and the decrease in energy obtained
by all other spheres at distances greater than $2^{1/6}$. Comparing DLP and
Lennard-Jones optimal configurations for the same number of points (sphere
centers), we find that sets of optimal packings only overlap for the trivial
cases $N=1$ and $N=2$; in general, they are completely different, with the
$V_{LJ}$ of the DLP optimal packings with optimized $D$ for $N=34$ to $N=161$
about $80\%\!-\!95\%$ of the minimal known $V_{LJ}$ for $N+1$ points. Figure
\ref{DLPLJ} depicts the minimal $V_{LJ}$ alongside the $V_{LJ}$ of DLP optimal
packings with optimized $D_{opt}(N)$ for $N=34$ to $N=161$.

In Fig. \ref{DLPLJ}, there are three $N$, $N=42$, $114$, and $134$, where the
$V_{LJ}$ for the DLP optimal packings and optimal configurations of $N+1$ points
with Lennard-Jones potential (\ref{LJenergy}) are particularly close
\cite{endnote1}. The proximity of the $V_{LJ}$ at these $N$ can be attributed to
the symmetry of the DLP optimal packings. It is known
\cite{Northby1987a,WD1997a} that minimal-energy configurations of points with
Lennard-Jones potential tend to favor icosahedral symmetry, either in part of
the configuration or in its entirety. The $N=42$ and $N=134$ DLP optimal
packings are also roughly icosahedrally symmetric, and though the $N=114$ DLP
optimal packing exhibits perfect three-fold rotational (chiral) symmetry, its
first and last shells are roughly icosahedrally symmetric.

Another minimal-energy problem involves finding the minimal-energy (optimal)
spatial configurations of $N+1$ identical nonoverlapping spheres where potential
energy $V_{sm}$ is defined in terms of the second moment about the centroid of
the $N+1$ sphere centers,
\begin{equation}
V_{sm}({\bf r}^{N+1}) = \sum_{i=1}^{N+1}|{\bf r}_i-{\bf C}|^2,
\label{SMenergy}
\end{equation}
with ${\bf C}\equiv (N+1)^{-1}\sum_{i=i}^{N+1}{\bf r}_i$ the centroid
\cite{SHDC1995a}. Comparing DLP and minimal second moment optimal packings up to
$N=32$, the largest $N$ for which minimal second moment optimal packings are
available, we find that only for $N=1$ to $N=4$ do sets of optimal packings
overlap, and as with the Lennard-Jones problem, in general they are completely
different.

The wide variance in optimal configurations at the same number of spheres
(points) across minimal-energy problems has implications for nucleation theory.
Comparing these three problems, we see that the functional form of the potential
energy has a substantial effect on the spatial arrangement of optimal
configurations, despite that the potentials in all three problems are isotropic
with long-range attractions and strong short-range repulsions. This suggests
that the sizes and shapes of critical nuclei in classical overcompressed liquids
(where dynamics are generally dominated by strong short-range repulsion) may
depend heavily on the precise functional form of the pair potential acting
between particles. The potential effects of the structures of dense nuclei on
the probability of freezing in overcompressed liquids will be discussed in
further detail in Sec. \ref{conclusions}.

\section{DLP optimal packings in three dimensions}
\label{thePackings}
Despite that the DLP optimal packings are almost always most similar, from among
the packings in ${\mathbb B}_N$, to a subset of an FCC Barlow packing or one of
its variants indistinguishable to the similarity metric, all optimal packings
with $N \geq 13$ are significantly more locally dense than any subset of a
Barlow packing including $N+1$ spheres. In general, over the range of $N$
studied in ${\mathbb R}^3$, we find wide variation in the symmetries, contact
networks, and other characteristics of DLP optimal packings, just as in
${\mathbb R}^2$.

For the majority of $N$, there are an uncountably infinite number (a continuum)
of DLP optimal packings with optimal radius $R_{min}(N)$, with the continuum
attributable to the presence of rattlers. A rattler in a packing of spheres in
${\mathbb R}^d$ is a sphere that is positioned such that it may be individually
moved in at least one direction without resulting overlap of any other sphere
within the packing or the packing boundary (in this case, the encompassing
sphere of radius $R_{min}(N) + 1/2$), i.e., a rattler is a sphere that is not
locally jammed \cite{TTD2000a,TS2001a}. Over the $184$ DLP optimal packings
studied between $N=34$ and $N=1054$, not including the central sphere, $170$
contain rattlers, with every packing for $N > 114$ containing at least one
rattler.

In the following figures (Figs. \ref{N13}-\ref{maracas}), only the backbones, or
the packings with the rattlers removed, are depicted, unless otherwise
specified. Additionally, each DLP packing is divided into shells, where a shell
in a DLP optimal packing is defined as all spheres with centers an equal
distance $R$ from the center of the central sphere. Each shell can be visualized
in the plane by employing a mapping to project points on a spherical surface in
${\mathbb R}^3$ (a shell) to a disk of radius $\pi$ in ${\mathbb R}^2$.
Considering a point in ${\mathbb R}^3$ in spherical coordinates and a point in
${\mathbb R}^2$ in polar coordinates, the mapping leaves the azimuthal angle
unchanged while the angle of inclination in ${\mathbb R}^3$ becomes the radius
in ${\mathbb R}^2$. The zenith direction from which the angle of inclination is
measured is generally selected to preserve angular symmetry. In Figs.
\ref{N13}-\ref{maracas}, points of distance unity (contacting spheres) in
${\mathbb R}^3$ are joined by lines.

\subsection{DLP optimal packings for $N\leq33$}
In ${\mathbb R}^d$ for any $d$, all DLP optimal packings with $N \leq K_d$ the
kissing number have $R_{min}(N) = 1$, with $K_3 = 12$ in ${\mathbb R}^3$. Also
for any $d$, the set of all DLP optimal packings for a given $N$ with
$R_{min}(N) \leq \tau$, $\tau = (1+\sqrt{5})/2$ the golden ratio, include
configurations where all $N$ sphere centers lie on a spherical surface of radius
$R_{min}(N) = R^S_{min}(N)$ \cite{HST2010a}. We recall that $R^S_{min}(N)$ is
the radius of the smallest spherical surface onto which the centers of $N$
spheres of unit diameter can be packed. The greatest $N$ for which $R_{min}(N)
\leq \tau$, denoted by $N_d^{\tau}$, in ${\mathbb R}^3$ is $N_3^{\tau} = 33$.
For $13 \leq N \leq 33$ in ${\mathbb R}^3$, our findings indicate that there are
no DLP optimal packings with jammed spheres with centers at a radius $r <
R_{min}(N)$.

\begin{figure}[!ht]
\centering
\subfigure[]{\includegraphics[width=2.1in,clip]{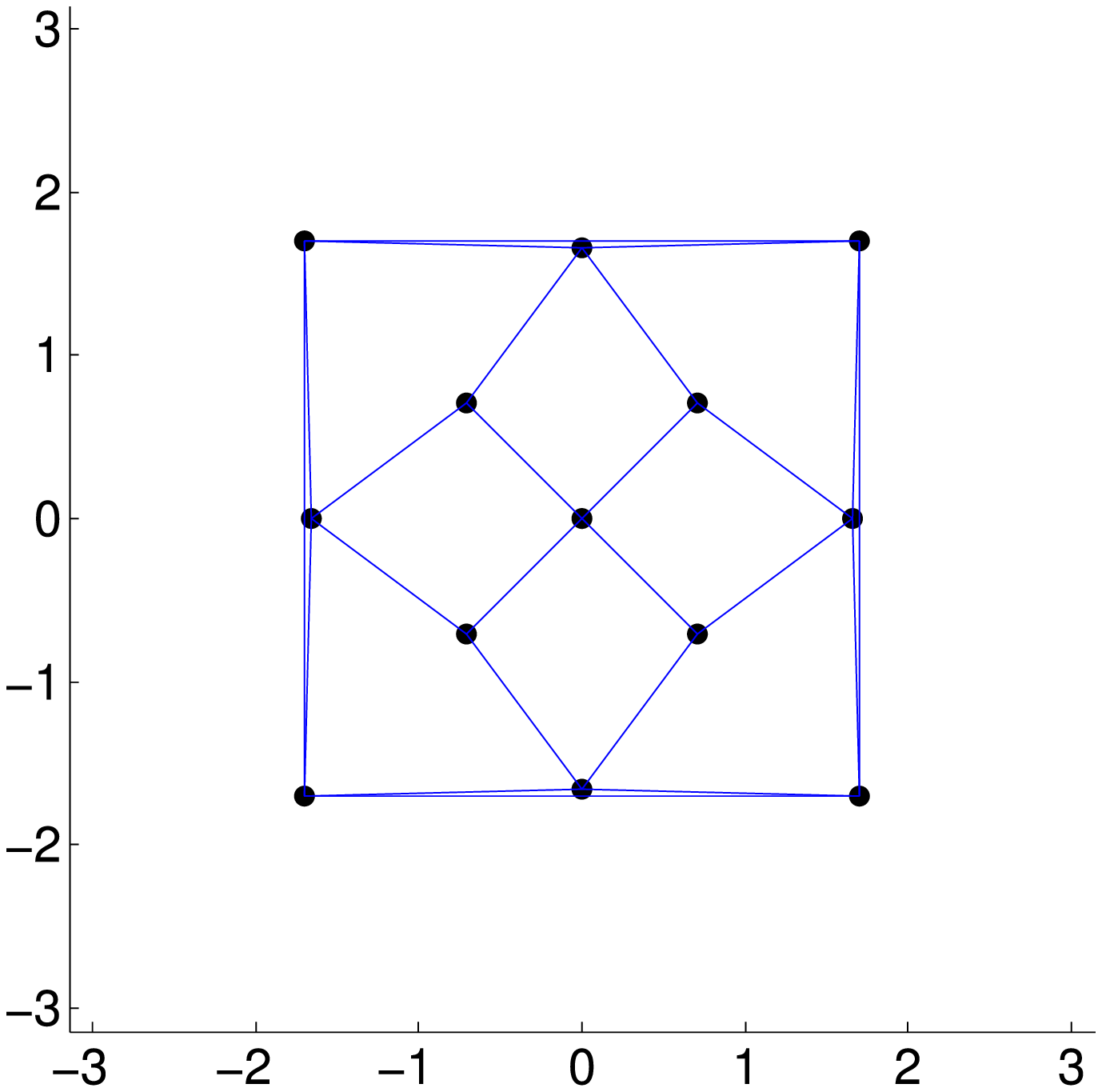}} \\
\caption{(Color online) The first and only shell of the $N=13$ DLP optimal
packing, which belongs to point group $C_{4v}$. This plot is a projection from
the surface of a sphere of radius $R_{min}(13) = 1.045573\dots$ of $13$ points
(sphere centers) in ${\mathbb R}^3$ to the interval $[-\pi,\pi] \times
[-\pi,\pi]$ in ${\mathbb R}^2$. In the projection, the angle of inclination of a
point in ${\mathbb R}^3$ represented in spherical coordinates becomes the
distance from the origin in ${\mathbb R}^2$, while the azimuthal angle remains
unchanged. Contacting spheres are connected by straight lines. In the above
projection, the zenith direction from which each point's angle of inclination is
measured is chosen to coincide with the packing's $C_4$ axis.}
\label{N13}
\end{figure}

Figure \ref{N13} is a projection to ${\mathbb R}^2$ of the $N=13$ DLP optimal
packing found by the algorithm. This configuration of spheres was first
documented in \cite{SW1951a}, where the authors conjectured that it was the
densest packing of $13$ nonoverlapping spheres of unit diameter with centers
restricted to a spherical surface of radius $R$. According to the principle
(proved in Ref. \cite{HST2010a}) that for $K_d < N \leq N_d^{\tau}$, $R_{min}(N)
= R^S_{min}(N)$, we conjecture that it is also the densest packing, without
restriction, of $13$ spheres around a fixed central sphere.

Unlike in ${\mathbb R}^2$, we know of no rigorous proofs indicating for $K_3 < N
\leq N_3^{\tau}$ what are the smallest radii $R$, equal to $R_{min}(N)$, onto
which the centers of $N$ identical nonoverlapping spheres of unit diameter may
be placed. However, we have found that the putative $R_{min}(N)$ found by the
algorithm for $13 \leq N \leq 33$ are equal to the strongly conjectured values
for $R^S_{min}(N)$ presented in \cite{SloaneSCweb}. This provides further
evidence that these smallest-known $R^S_{min}(N) = R_{min}(N)$ are optimal.

\subsection{Particularly dense, symmetric optimal packings}
\label{denseSymm}
For four values of $N$, $N=60,62,84$ and $114$, a highly symmetric arrangement
of spheres yields a packing that is significantly more locally dense than DLP
optimal packings of nearby $N$. The relatively high densities of these packings
appears in Fig. \ref{ZmaxVsZBar} as upturns above the linear trend and in Fig.
\ref{RminVsRBar} as downturns below the one-third power (in $N$) trend. All $N$
surrounding spheres in each of these four packings are locally jammed, though in
two of the
four cases, the central sphere is not in contact with any of its nearest
neighbors.

\begin{figure}[!ht]
\centering
\subfigure[]{\includegraphics[width=2.1in,clip]{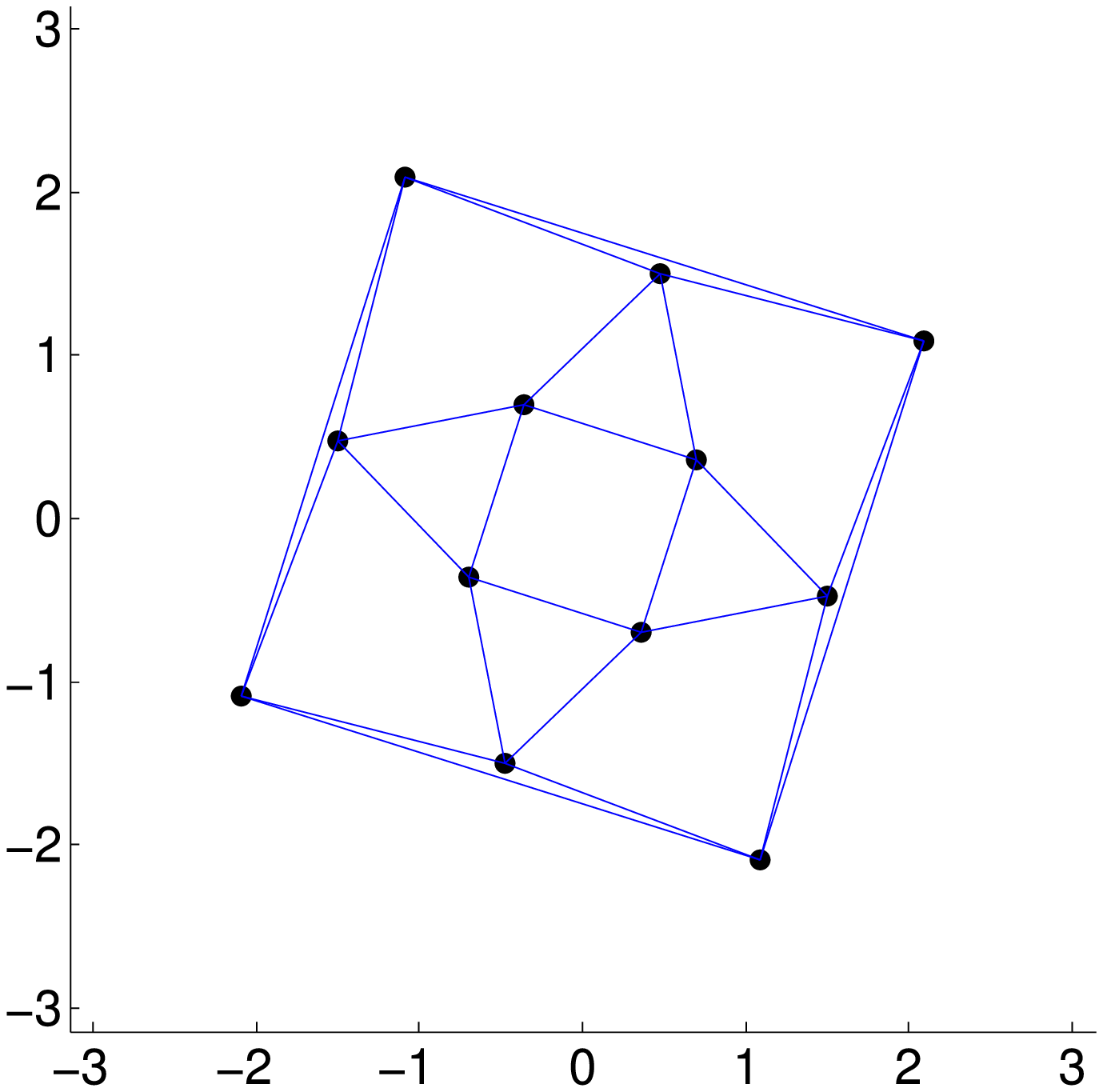}} \\
\subfigure[]{\includegraphics[width=2.1in,clip]{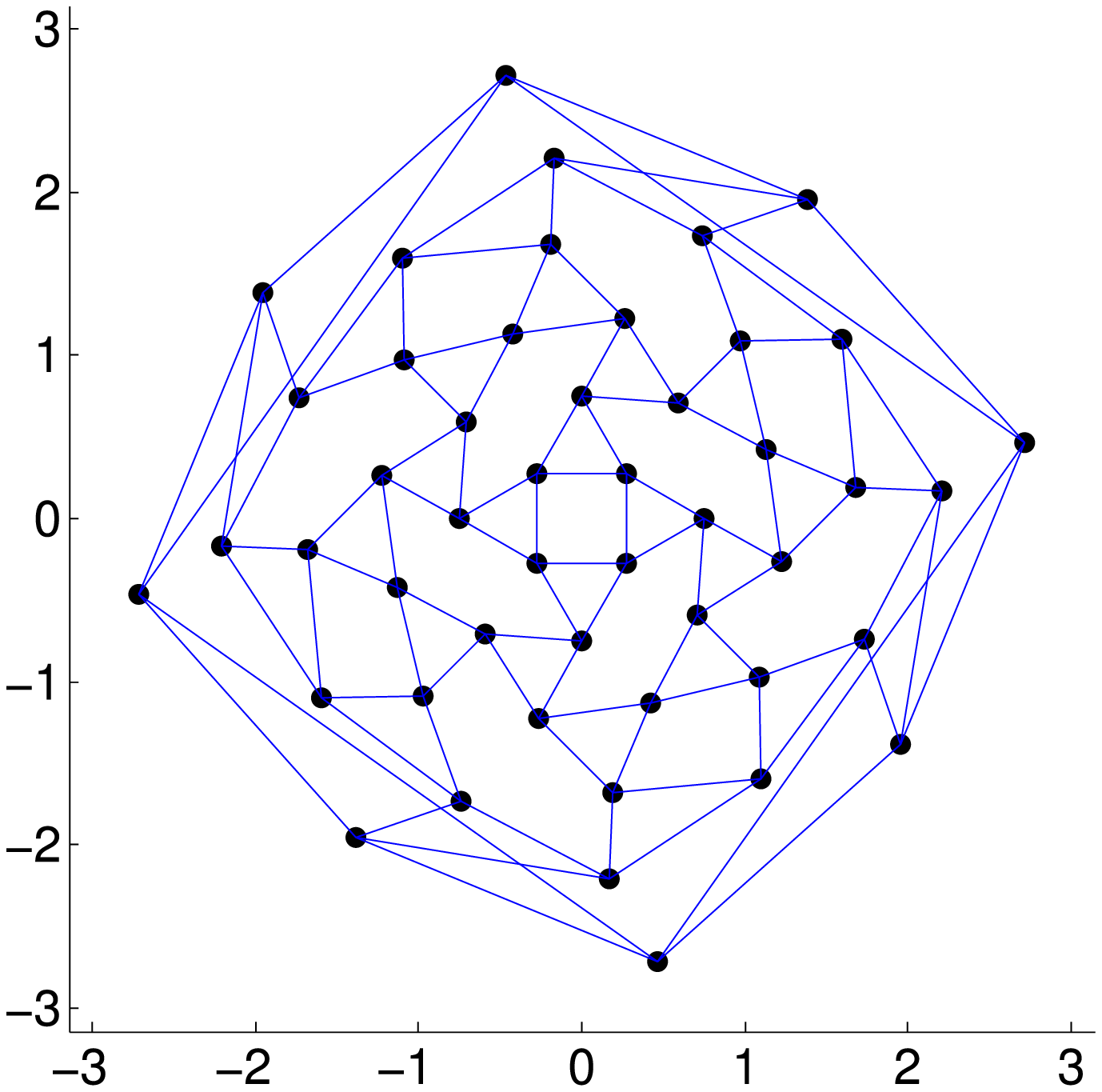}}
\caption{(Color online) The two shells of the $N=60$ DLP optimal packing, which
belongs to point group $O$. The zenith direction is chosen along one of the
$C_4$ axes shared by both shells. (a) $12$ spheres, $R = 1$, point group $O_h$.
(b) $48$ spheres, $R = R_{min}(60) = 1.891101\dots$, point group $O$.}
\label{N60}
\end{figure}

\begin{figure}[!ht]
\centering
\subfigure[]{\includegraphics[width=2.1in,clip]{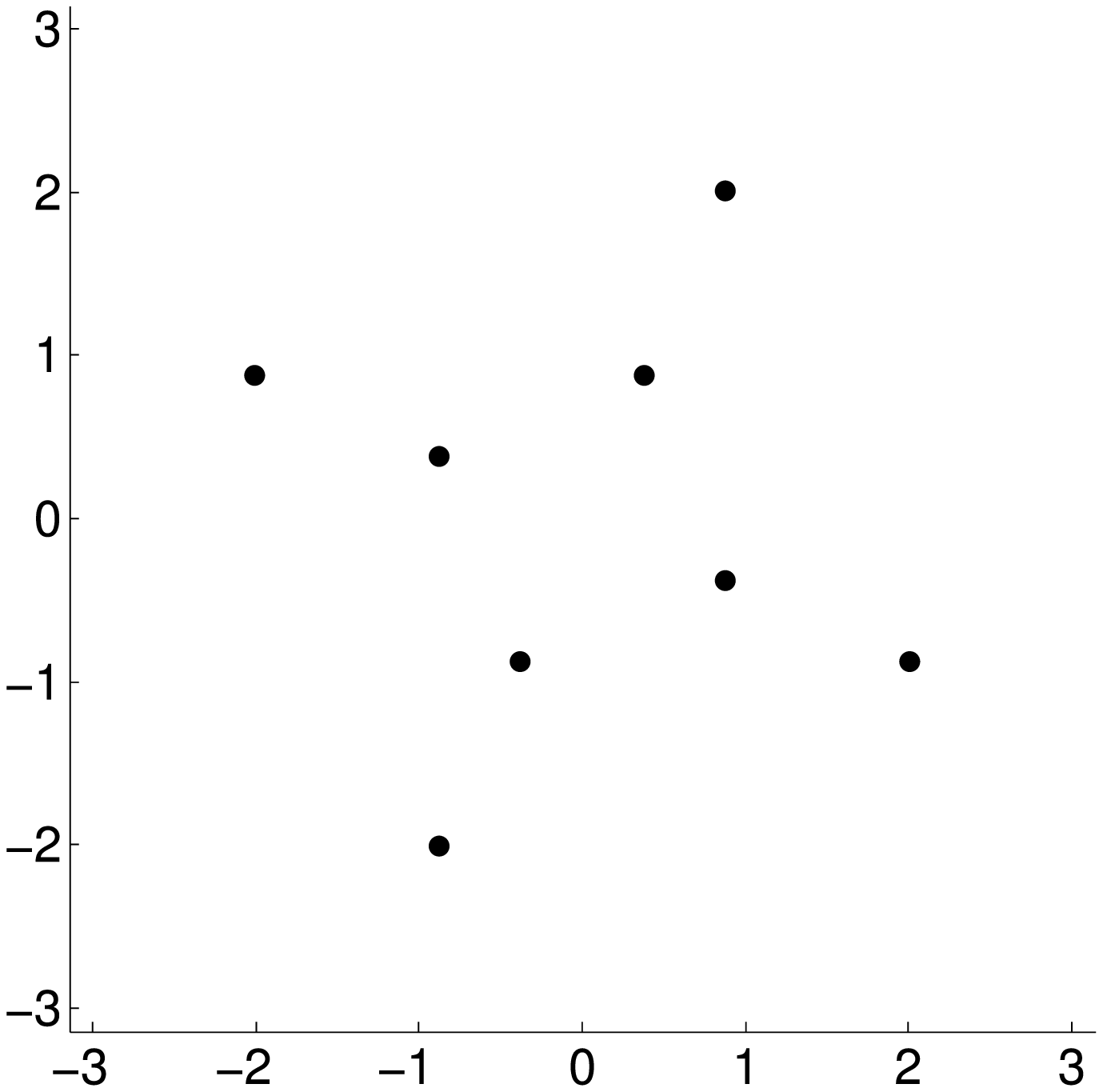}} \\
\subfigure[]{\includegraphics[width=2.1in,clip]{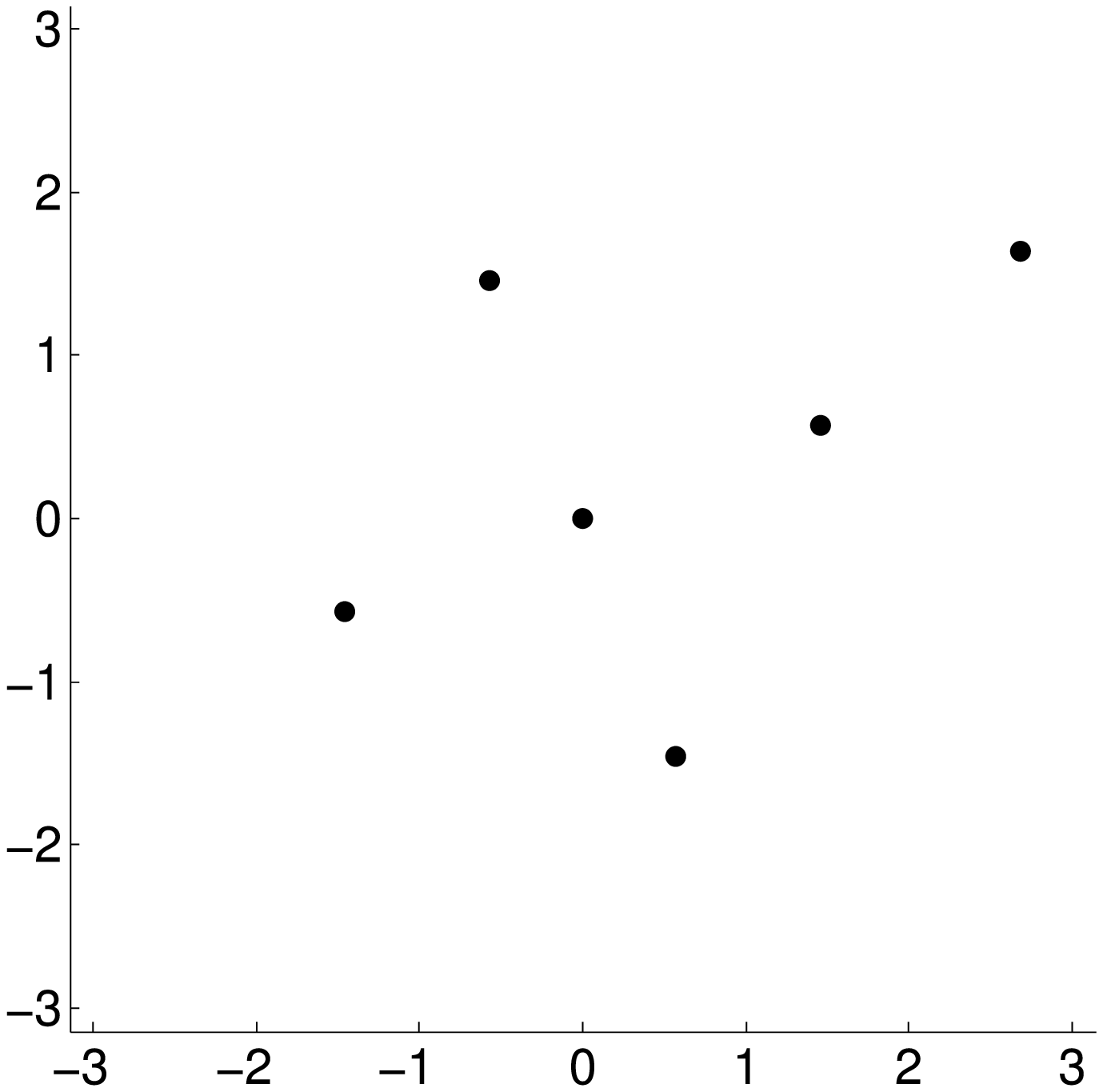}} \\
\subfigure[]{\includegraphics[width=2.1in,clip]{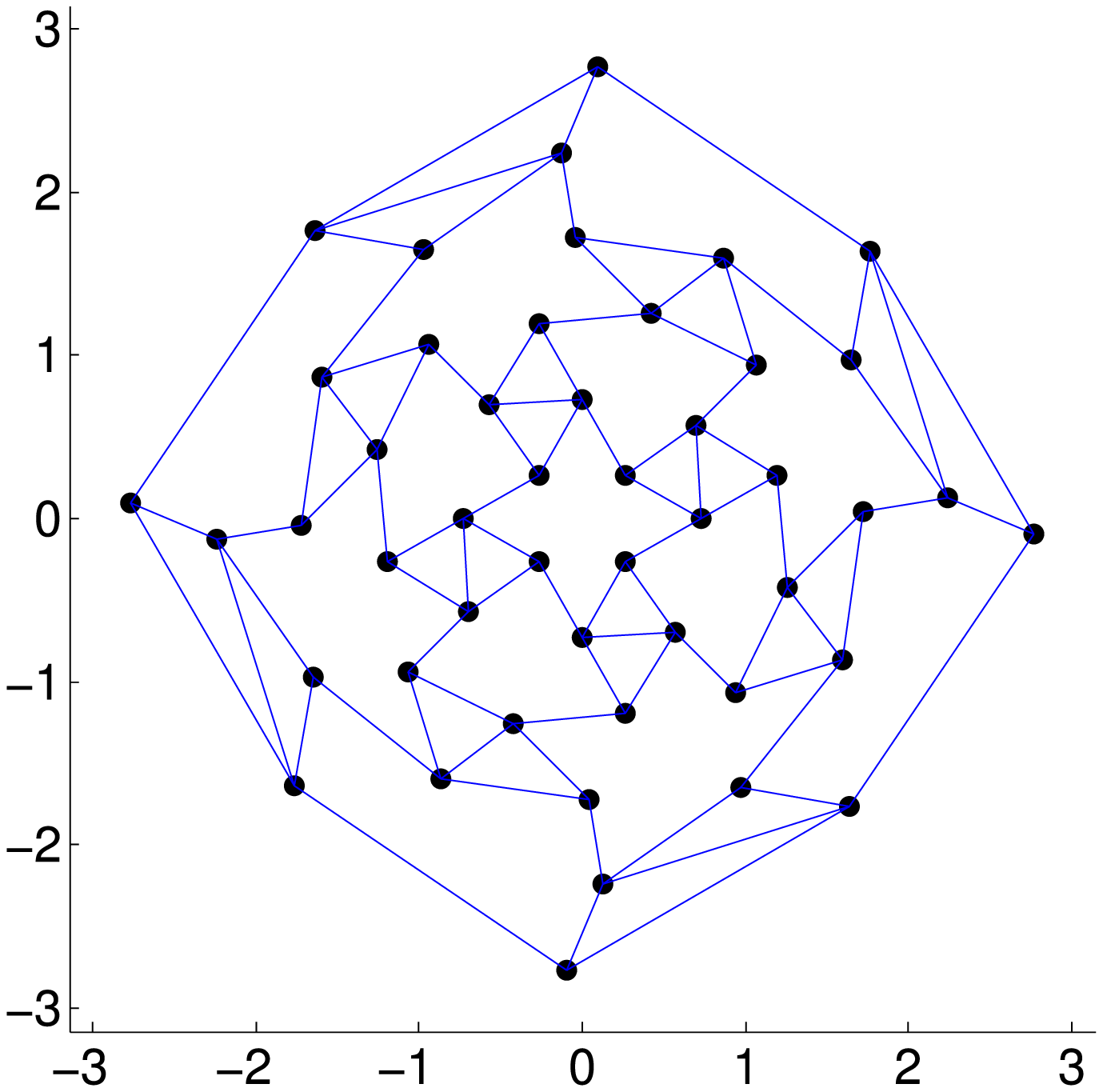}}
\caption{(Color online) The three shells of the $N=62$ DLP optimal packing,
which belongs to point group $O$. The zenith direction is chosen along one of
the $C_4$ axes shared by all three shells. (a) $8$ spheres, $R = 1.087542\dots$,
point group $O_h$. (b) $6$ spheres, $R=1.087786\dots$, point group $O_h$. In
this projection, as the sphere center furthest from the origin is at distance
$\pi$ (corresponding in ${\mathbb R}^3$ to an angle of inclination equal to
$\pi$), its azimuthal angle is chosen at random. (c) $R = R_{min}(62) =
1.927716\dots$, point group $O$.}
\label{N62}
\end{figure}

\begin{figure}[!ht]
\centering
\subfigure[]{\includegraphics[width=2.1in,clip]{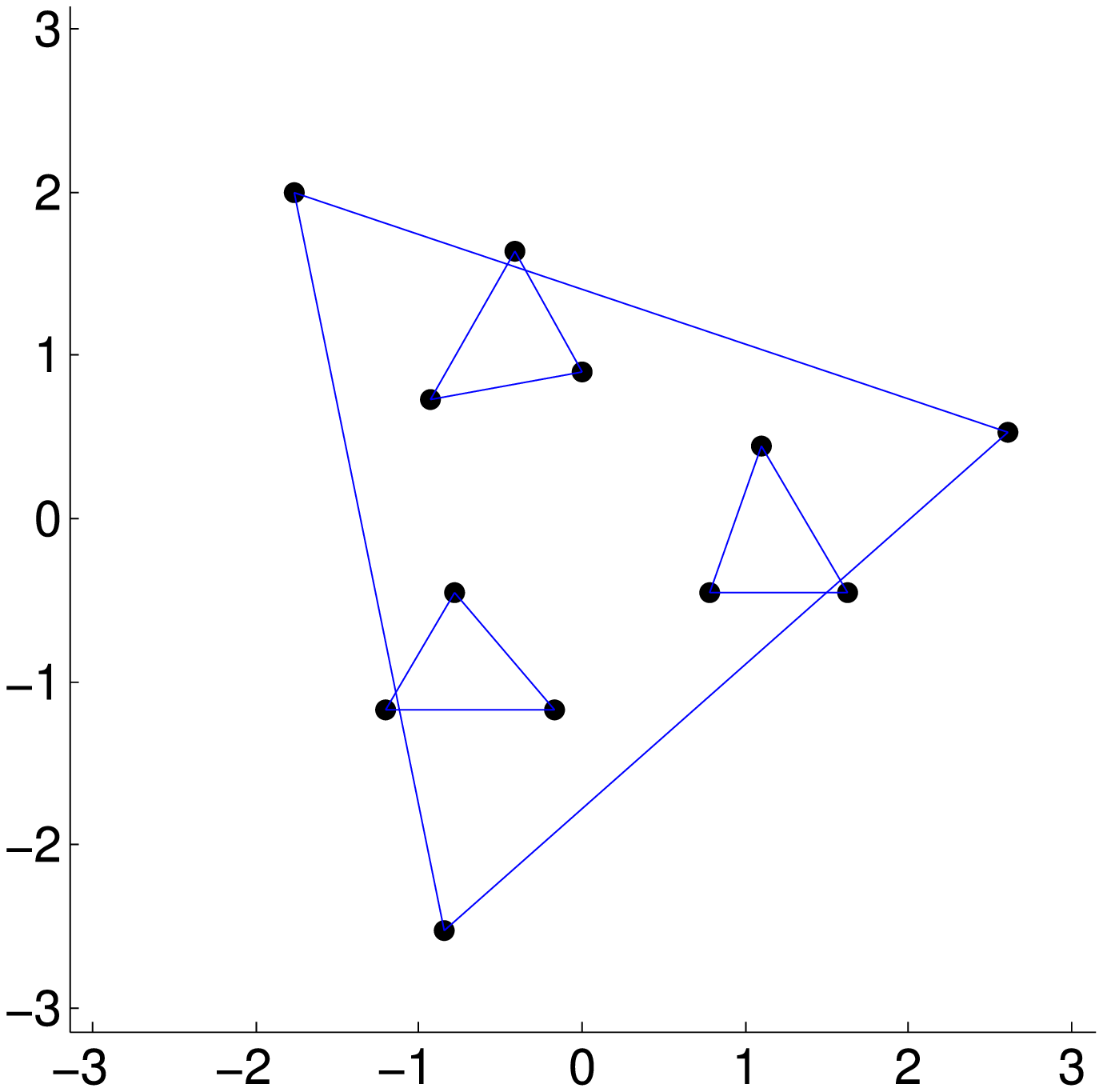}} \\
\subfigure[]{\includegraphics[width=2.1in,clip]{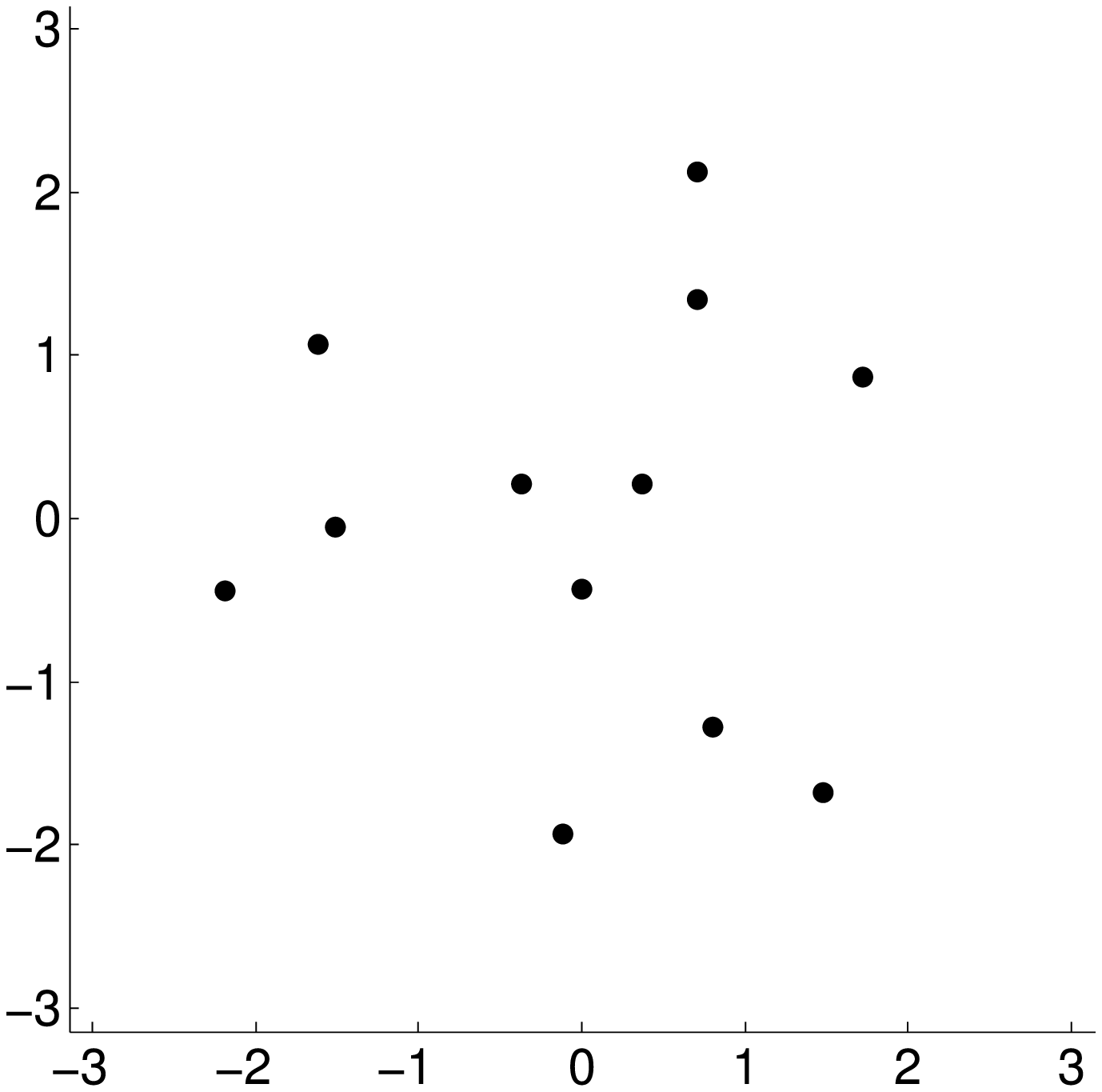}} \\
\subfigure[]{\includegraphics[width=2.1in,clip]{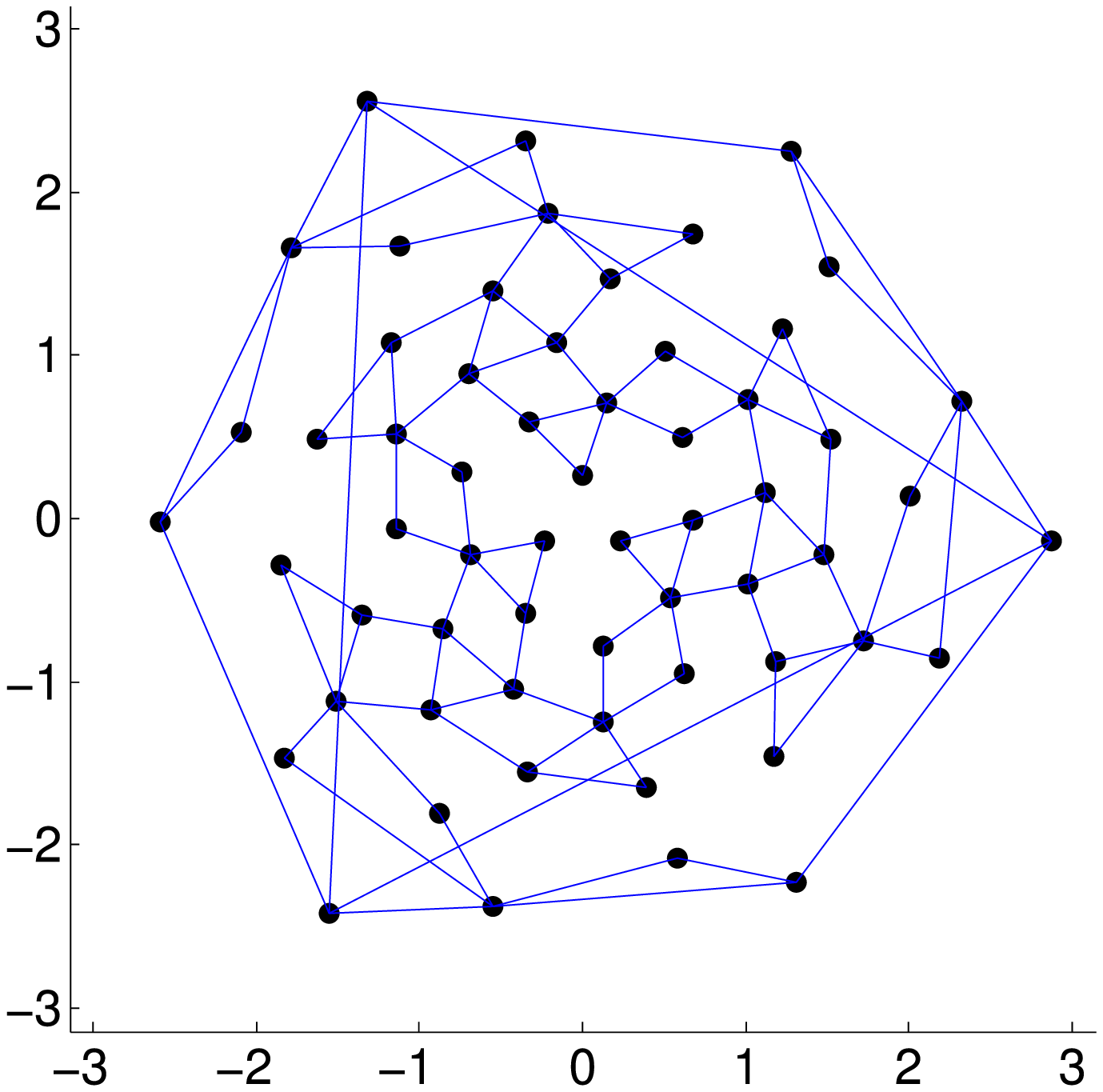}}
\caption{(Color online) The three shells of the $N=84$ DLP optimal packing,
which belongs to point group $T$. The zenith direction is chosen along one of
the $C_3$ axes shared by all three shells. (a) $12$ spheres, $R =
1.255451\dots$, point group $T$. (b) $12$ spheres, $R = 1.423714\dots$, point
group $T$. (c) $60$ spheres, $R = R_{min}(84) = 2.182390\dots$, point group
$T$.}
\label{N84}
\end{figure}

Figure \ref{N60} depicts the two shells of the $N=60$ DLP optimal packing,
which has rotational (chiral) octahedral symmetry. The first shell contains
$12$ spheres (the zeroeth shell is the central sphere) arranged with full
octahedral symmetry and in contact with the central sphere. The second shell
contains $48$ spheres with rotational octahedral symmetry and with centers at
distance $R_{min}(60) = 1.891101\dots$. It is of note that the $N=60$ optimal
packing is such a relatively densely-packed configuration of spheres that
$R_{min}(59) = R_{min}(60)$, and DLP optimal packings for $N=59$ can be formed
simply by deleting any one of the $60$ surrounding spheres in the $N=60$
packing.

\begin{figure}[!ht]
\centering
\subfigure[]{\includegraphics[width=2.1in,clip]{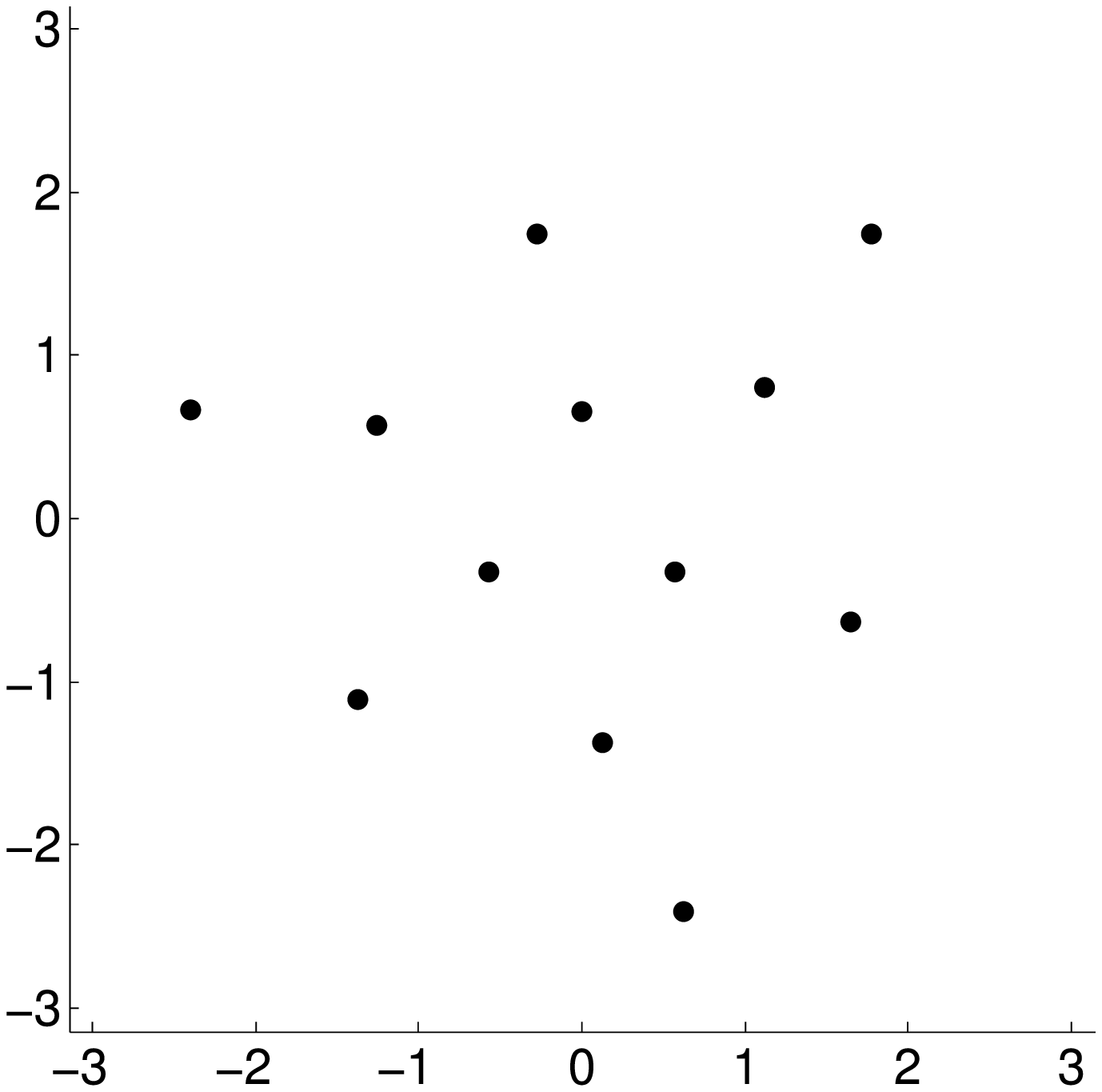}} \\
\subfigure[]{\includegraphics[width=2.1in,clip]{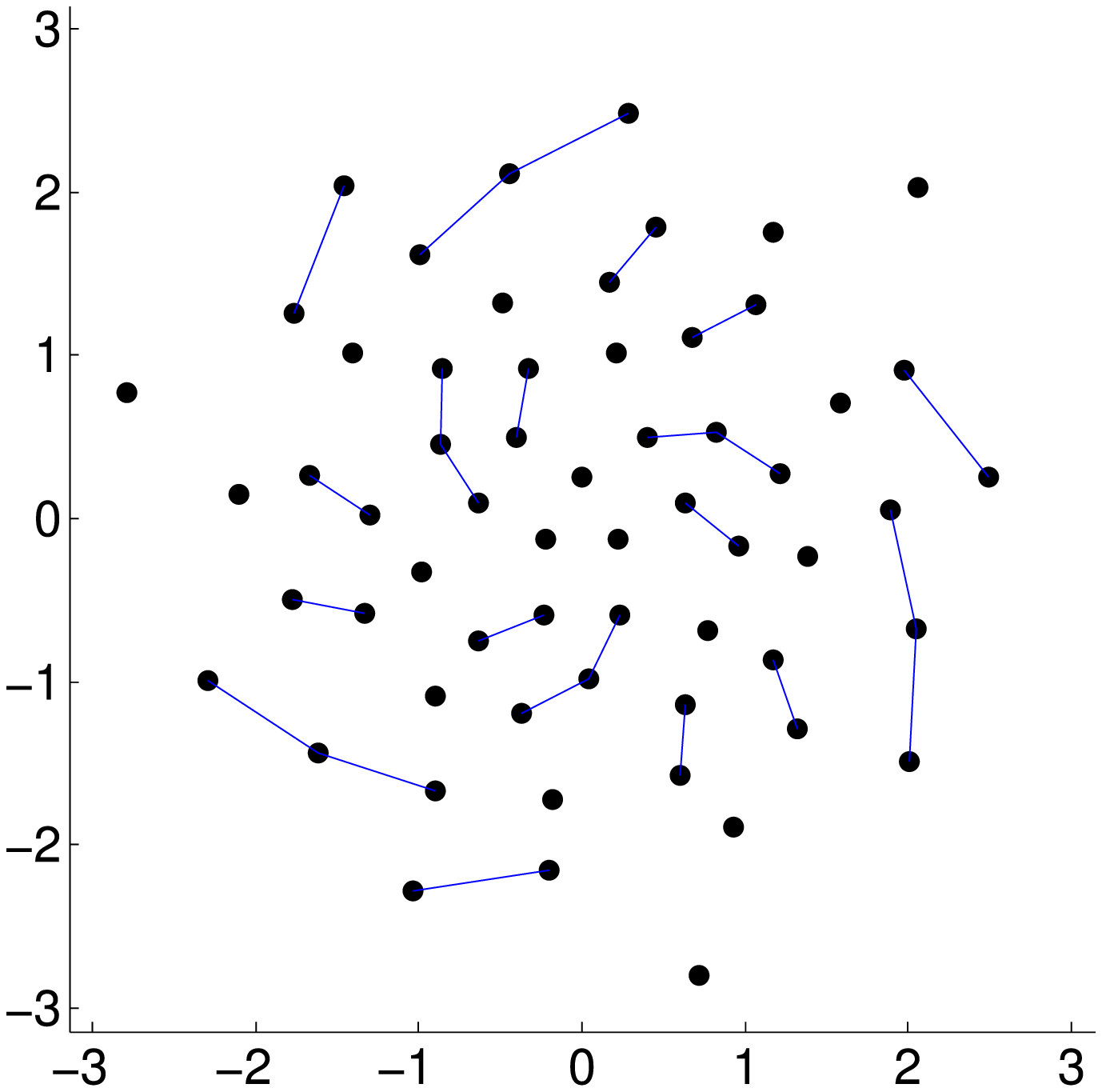}}
\caption{(Color online) The first and tenth shells of the $N=114$ DLP optimal
packing, which has ten shells and belongs to point group $D_3$. The zenith
direction is chosen along one of the $C_3$ axes shared by all ten shells. (a)
$12$ spheres, $R = 1$, point group $D_3$. (b) $60$ spheres, $R = R_{min}(114) =
2.456227\dots$, point group $D_3$.}
\label{N114}
\end{figure}

Figure \ref{N62} depicts the three shells of the $N=62$ DLP optimal packing. The
shells contain $8$, $6$, and $48$ spheres, respectively, at distances $R =
1.087542\dots$, $R = 1.087786\dots$ and $R_{min}(62) = 1.927716\dots$ from the
center of the central sphere. The first and second shell have full octahedral
symmetry, while the third and the packing as a whole exhibit only rotational
octahedral symmetry. The first and second layers of the $N=62$ packing are
radially less than $2.5\times 10^{-4}$ sphere diameters from one another, and
together they form a cavity within which the central sphere, were it not fixed,
could move. This effect also occurs with disks in ${\mathbb R}^2$, as is
discussed in paper I \cite{endnote3}. More detail on this topic is presented in
Sec. \ref{cavity}.

Figure \ref{N84} shows the three shells of the $N=84$ DLP optimal packing, all
of which have rotational tetrahedral symmetry. The shells contain $12$, $12$,
and $60$ spheres, respectively, at distances $R = 1.255451\dots$, $R =
1.423714\dots$, and $R_{min}(84) = 2.182390\dots$ from the center of the central
sphere. The $N=84$ DLP optimal packing is unique in that it is the only optimal
packing that we have found, in ${\mathbb R}^2$ or ${\mathbb R}^3$, that exhibits
perfect tetrahedral symmetry. 

The $N = 114$ DLP optimal packing is composed of ten shells. The first
includes $12$ spheres in contact with the central sphere, and the tenth $60$
spheres at $R_{min}(114) = 2.456227\dots$; both have chiral three-fold dihedral
symmetry. Shells two and six each contain three spheres with centers arranged as
an equilateral triangle (point group $D_{3h}$), and shells three through five
and seven through nine each contain six spheres arranged with chiral three-fold
dihedral symmetry. Shells two through nine can be grouped into four pairs based
on radial distance from the central sphere. Each pair is no more than
$2.22\times 10^{-3}$ sphere diameters apart (shells two and three), but at least
$1.24\times 10^{-6}$ sphere diameters apart (shells eight and nine). Figure
\ref{N114} is an image of shells one and ten.

\subsection{Other optimal packings with high symmetry}
\label{otherSymm}
Over the $N$ studied, a large number of DLP optimal packings were found, aside
from those discussed in Sec. \ref{denseSymm}, that exhibit perfect symmetry. A
representative selection of these packings is presented in this section.

\begin{figure}[ht]
\centering
\subfigure[]{\includegraphics[width=2.1in,clip]{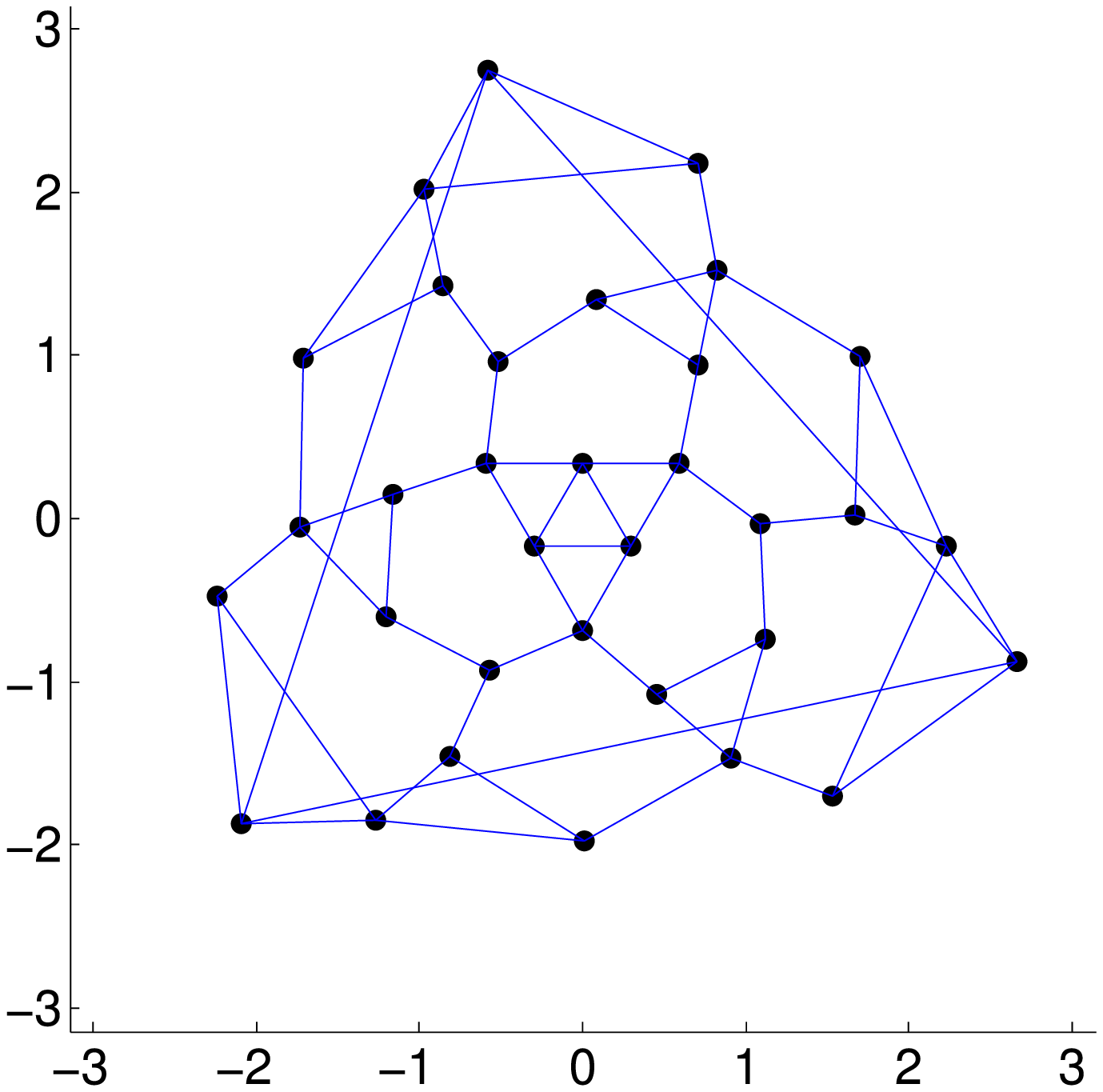}} \\
\subfigure[]{\includegraphics[width=2.1in,clip]{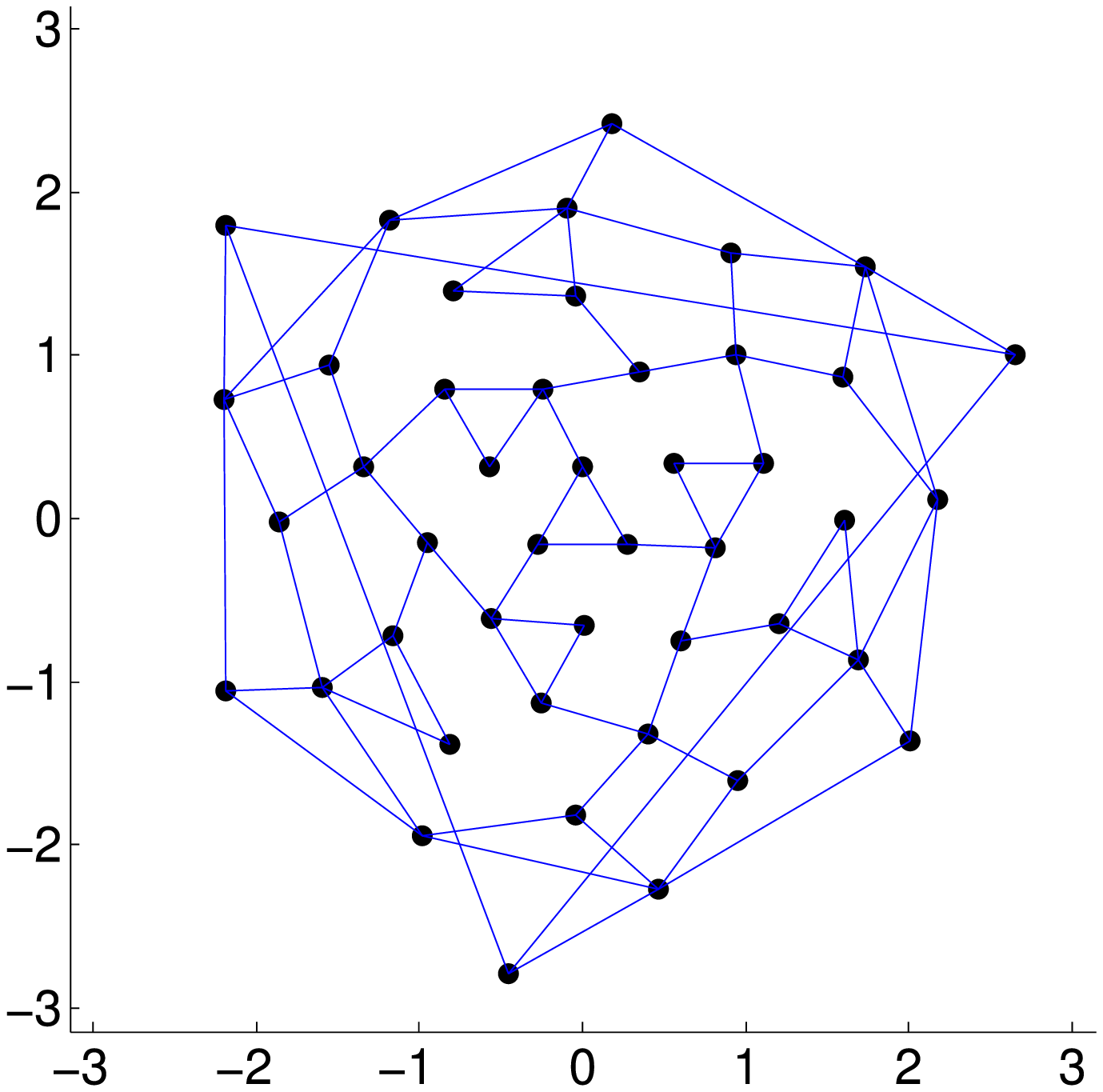}}
\caption{(Color online) The last shells of the $N=45$ and $N=57$ DLP optimal
packings, both of which belong to point group $C_3$. For both shells, the zenith
direction is chosen along one of the $C_3$ axes shared by all shells in the
packing. (a) $33$ spheres, $R = R_{min}(45) = 1.749670\dots$, point group $C_3$.
(b) $45$ spheres, $R = R_{min}(57) = 1.877196\dots$, point group $C_3$.}
\label{threeFold}
\end{figure}

Figure \ref{threeFold} depicts the last shells in the optimal packings for
$N=45$ and $N=57$ spheres; both packings have three-fold cyclic symmetry (point
group $C_3$). Also in both packings, all spheres including the central sphere
are jammed. The $N=45$ packing has four shells containing $6$, $3$, $3$, and
$33$ spheres with centers at radial distance $R=1$, $R=1.005960\dots$, $R =
1.032049\dots$, while $R_{min}(45) = 1.749670\dots$, respectively. The $N=57$
packing has three shells containing $9$, $3$, and $45$ spheres with centers at
radial distance $R=1$, $R=1.009196\dots$, and $R_{min}(57) = 1.877196\dots$,
respectively. 

\begin{figure}[ht]
\centering
\subfigure[]{\includegraphics[width=2.1in,clip]{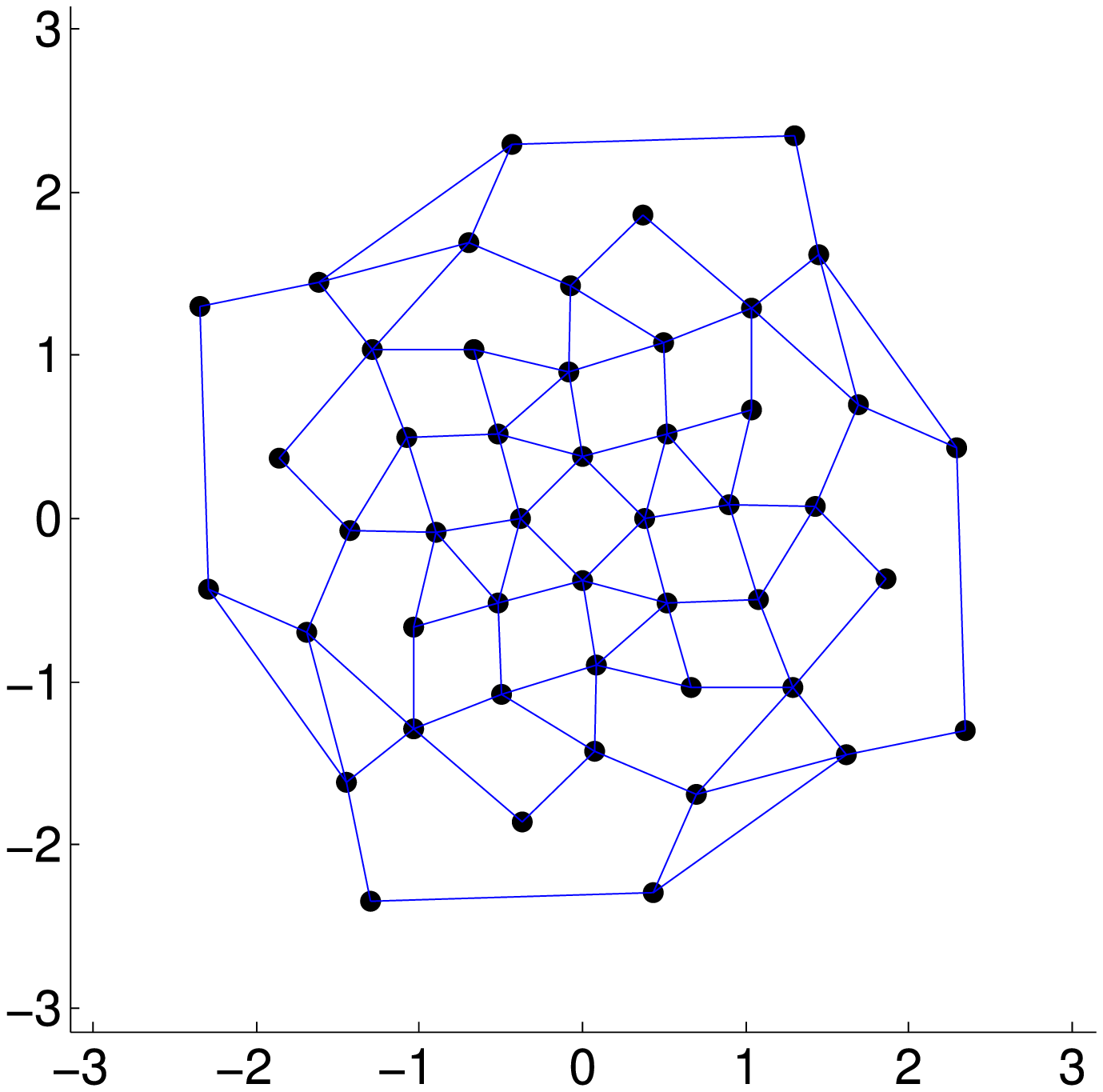}} \\
\subfigure[]{\includegraphics[width=2.1in,clip]{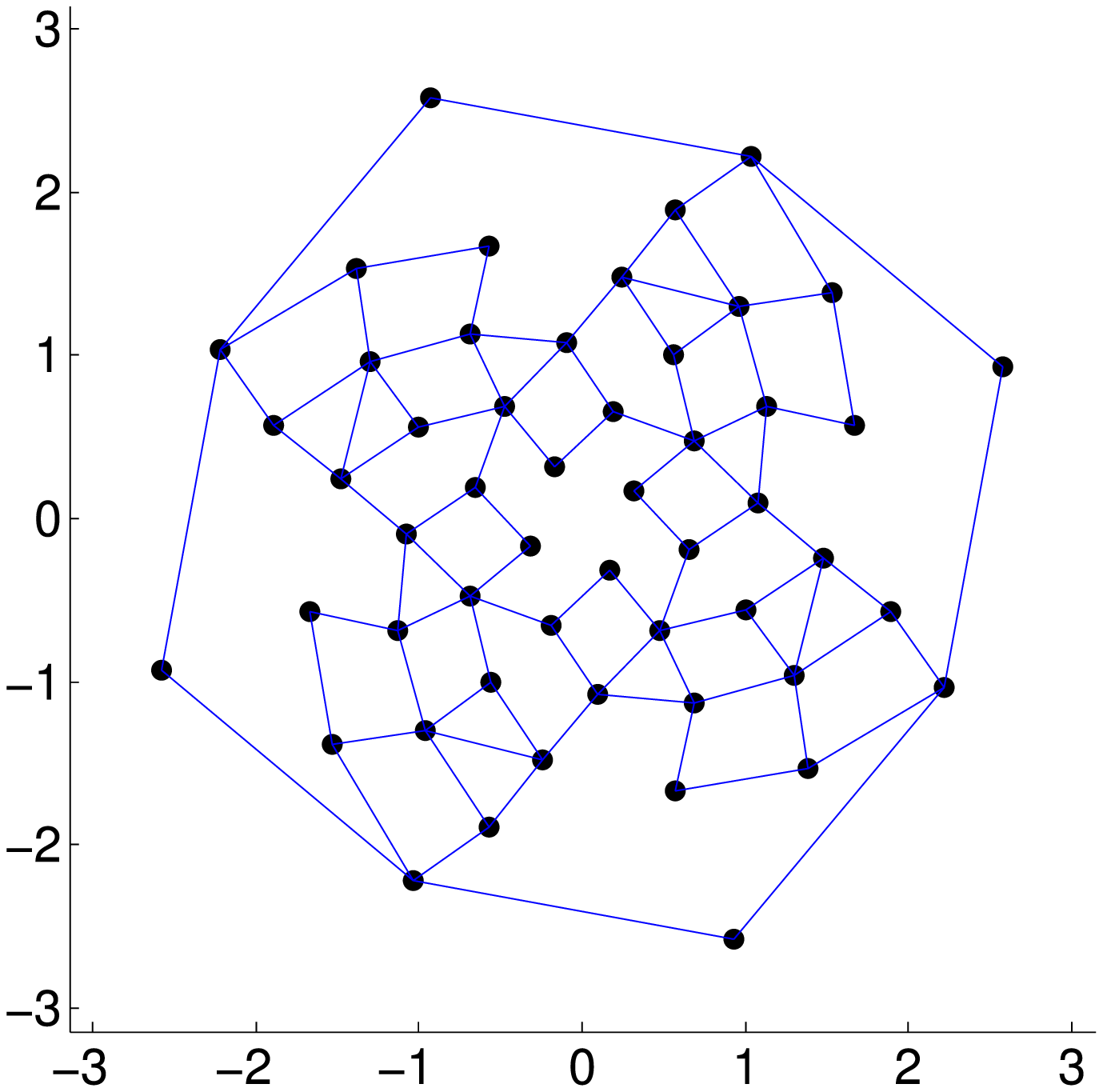}}
\caption{(Color online) The jammed spheres in the last shells of the $N=61$ and
$N=74$ DLP optimal packings, both of which belong to point group $C_4$. For both
shells, the zenith direction is chosen along one of the $C_4$ axes. (a) $48$
spheres, $R = R_{min}(61) = 1.919927\dots$, point group $C_4$. (b) $52$ jammed
spheres (of $56$ total in the shell), $R = R_{min}(74) = 2.077792\dots$, point
group $C_4$.}
\label{fourFold}
\end{figure}

Neither the $N=61$ nor the $N=74$ optimal packings are perfectly symmetric; both
belong to point group $C_1$. However, with one sphere removed, the remaining
$61$ spheres of the $N=61$ packing have four-fold cyclic symmetry (point group
$C_4$), and with two jammed spheres and four rattlers (from the last shell)
removed, the remaining $69$ spheres of the $N=74$ packing also have four-fold
cyclic symmetry. Figure \ref{fourFold} is an image of the jammed spheres in the
last shells in the $N=61$ and $N=74$ DLP optimal packings. Neither packing has a
jammed central sphere.

\begin{figure}[!ht]
\centering
\includegraphics[width=2.1in,clip]{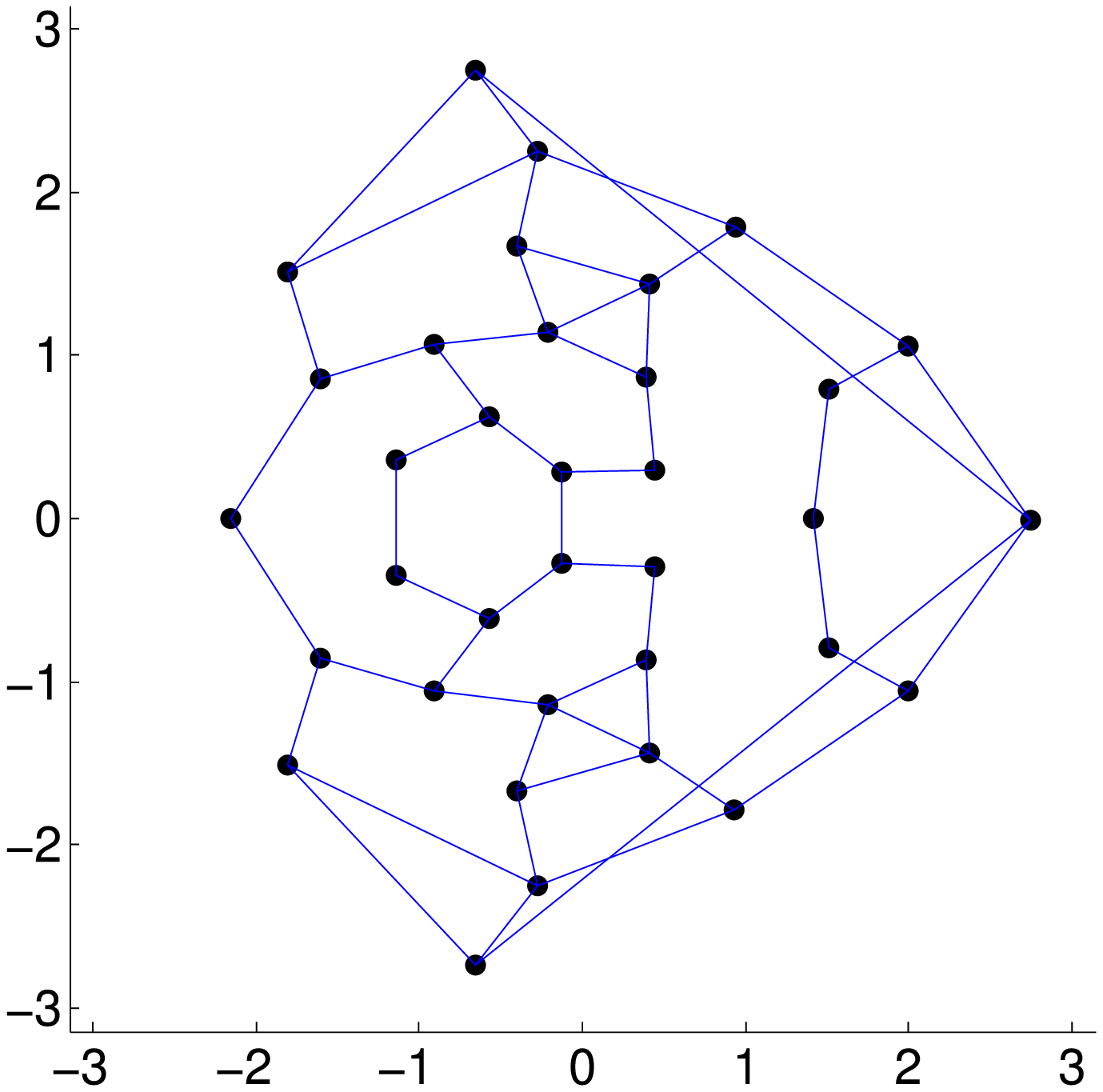}
\caption{(Color online) The last shell of the $N=50$ DLP optimal packing, which
contains $35$ jammed spheres with centers at distance $R_{min}(50) =
1.814049\dots$ and belongs to point group $C_s$. The zenith direction in the
image is chosen parallel to the mirror plane and through the contact point of
two spheres.}
\label{twoFold}
\end{figure}

Figure \ref{twoFold} is an image of the eighth and last shell of the $N=50$
DLP optimal packing; the shell contains $35$ jammed spheres with centers at
radial distance $R_{min}(50) = 1.814049\dots$ from the center of the central
sphere. The first three shells form a cavity around the central sphere and
contain four, two, and two spheres with centers at radial distances $R = 1$,
$R=1.000608\dots$, and $R=1.037107\dots$, respectively. The fourth through
seventh shells are single spheres positioned at distances $1.108 < R < 1.151$,
and the $35$ jammed spheres in the eighth shell are arranged with mirror
reflection symmetry (point group $C_s$). The $N=50$ packing is a particularly
good example of the complicated structures resulting from many-bodied
interactions in DLP optimal packings, in that it is not symmetric as a whole
(point group $C_1$) but contains shells exhibiting perfect symmetry that does
not appear related to the geometry of a spherical surface. It is also
interesting to note that the eighth shell of the $N=50$ DLP optimal packing is
the {\it only} example for $N\geq34$ in ${\mathbb R}^3$ of an achiral last
shell, i.e., it is the only last shell exhibiting a plane of reflection
symmetry.

\subsection{Imperfect icosahedral symmetry in DLP optimal packings}
\label{icosPackings}
Perhaps equally as interesting as the DLP optimal packings in ${\mathbb R}^3$
that are perfectly symmetric are those packings that exhibit only rough or
imperfect symmetry. For example, no single shell in a DLP optimal packing over
the $N$ studied exhibits perfect five-fold rotational or higher symmetry; in
particular, no shell exhibits perfect icosahedral symmetry. This is not the case
in ${\mathbb R}^2$, where there are three DLP optimal packings (for $N=10$,
$N=15$ and $N=25$) with perfect five-fold rotational symmetry, and a large
number with perfect six-fold rotational symmetry.

In ${\mathbb R}^2$, the kissing number $K_2 = 6$, and there is only one way (up
to rotations) to arrange six identical disks in contact with a same-size central
disk: each of the six must contact two of the remaining five, such that the
arrangement enclosed in an encompassing disk is jammed. It is therefore not
particularly surprising that six-fold rotational symmetry is common in DLP
optimal packings in ${\mathbb R}^2$. In ${\mathbb R}^3$, the kissing number $K_3
= 12$, but there is an infinite number of ways to arrange $12$ identical spheres
in contact with a same-size central sphere. 

The differences between configurations of $K_d$ spheres in contact with the
central sphere in dimensions two and three can serve as an explanation for why
perfect icosahedral symmetry is not found, over the $N$ studied, in the DLP
optimal packings in ${\mathbb R}^3$. In ${\mathbb R}^3$, the central sphere
contributes to the disruption of perfect icosahedral symmetry by preventing the
$12$ surrounding spheres from forming an icosahedron of contacting spheres, as
$12$ identical contacting spheres with centers on the vertices of an icosahedron
is the densest packing of $12$ spheres around a point.

Supporting this explanation, imperfect icosahedral symmetry is present in a
significant number of the first shells in DLP optimal packings. Specifically, if
the tolerance for sphere center overlap used in calculating symmetry elements is
raised from $10^{-8}$ to $0.2$ sphere diameters, then the $12$ spheres in
contact with the central sphere in the first shells of the DLP optimal packings
for the $N=42$, $N=114$, $116$, $117$, $118$, $133$, $135\!-\!139$, and
$530\!-\!532$ DLP optimal packings all exhibit icosahedral symmetry.
Additionally, the $12$ spheres closest to the central sphere in the $N=269$,
$320$, $533$, and $886$ packings exhibit icosahedral symmetry within a tolerance
of $0.2$ sphere diameters.

Imperfect icosahedral symmetry is present in many DLP optimal packings at radial
distances much greater than unity. In general, for $N$ near $N=42$ and $N=134$,
DLP optimal packings exhibit imperfect icosahedral symmetry throughout the
entire packing. This is due to the existence of two particularly dense,
perfectly icosahedrally symmetric packings containing two and four shells,
respectively, totaling $42$ and $134$ identical nonoverlapping spheres
surrounding a central same-size sphere. The first $134$ spheres in the $N=269$,
$320$, and $530\!-\!533$ DLP optimal packings can be described as variations on
the $134$-sphere packing, which contains as a subset the $42$-sphere packing.
Table \ref{icosTable} provides the details of the polyhedra composing the
perfectly symmetric $134$-sphere packing, where the fourth shell of the packing
includes $60$ spheres with centers on the vertices of a truncated icosahedron
and an additional $20$ spheres with centers arranged along radial vectors from
the center of the packing through the centers of each of the truncated
icosahedron's $20$ regular hexagonal faces.

\begin{table}
\centering
\caption{Details of a dense, perfectly icosahedrally symmetric packing of $134$
identical nonoverlapping spheres around a same-size central sphere.}
\begin{tabular}{c|c c c c c}
\hline
\hline
& & & Side & \,\,Exact vertex\,\, & Num. vertex \\
\,\,Shell \,\,&\,\, Spheres & Shape & length & distance & distance \\
\hline
\,\,$1$\,\, & \,\,$12$ & icosahedron & $\frac{2}{(\tau+2)^{1/2}}$ & $1$ &
$1$\,\, \\
\,\,$2$\,\, & \,\,$30$ & icosidodecahedron & $\frac{2}{(\tau+2)^{1/2}}$ &
$\frac{2\tau}{(\tau+2)^{1/2}}$ & $1.701302\dots$ \\
\,\,$3$\,\, & \,\,$12$ & icosahedron & $\frac{4}{(\tau+2)^{1/2}}$ & $2$ &
$2$\,\, \\
\,\,$4$\,\, & \,\,$80$ &\,\, truncated icosahedron\,\, &
$\frac{2}{(\tau+2)^{1/2}}$ & $\left(\frac{9\tau+10}{\tau+2}\right)^{1/2}$ &
$2.605543\dots$ \\
\hline
\end{tabular}
\label{icosTable}
\end{table}

The numbers of spheres in the outer shells of the two perfectly icosahedrally
symmetric packings are near to the maximal number $Z^S_{max}(R_{min}(N))$ that
can be placed on spherical surfaces of radii $R_{min}(42)$ and $R_{min}(134)$,
respectively. As a result, DLP optimal packings at $N$ near $42$ and $134$ can
be packed similarly to these icosahedrally symmetric packings while roughly
adhering to the empirical rule of surface-maximization. However, none of the
spheres in any of the perfectly icosahedrally symmetric packing's shells are in
contact with any other spheres within that shell, and the spheres in the second
icosahedron (third shell) are not in contact with any of the spheres in the
icosidodecahedron (second shell). This lack of contact allows the spheres to be
translated away from their perfectly icosahedrally symmetric positions in order
to obtain a smaller DLP optimal packing radius $R_{min}(N)$ for $N$ near $N=42$
and $N=134$.

\begin{figure}[!ht]
\centering
\subfigure[]{\includegraphics[width=2.1in,clip]{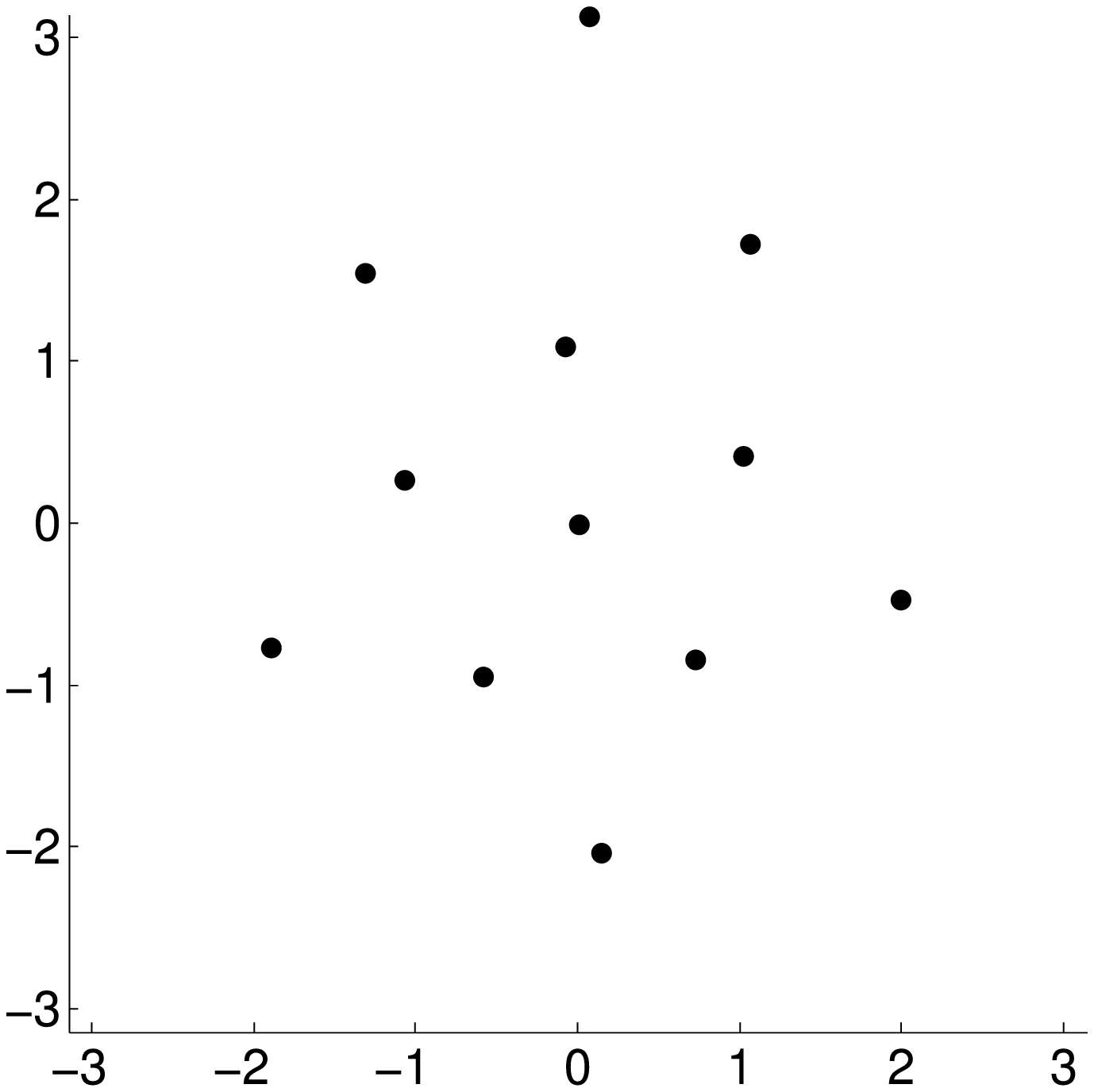}} \\
\subfigure[]{\includegraphics[width=2.1in,clip]{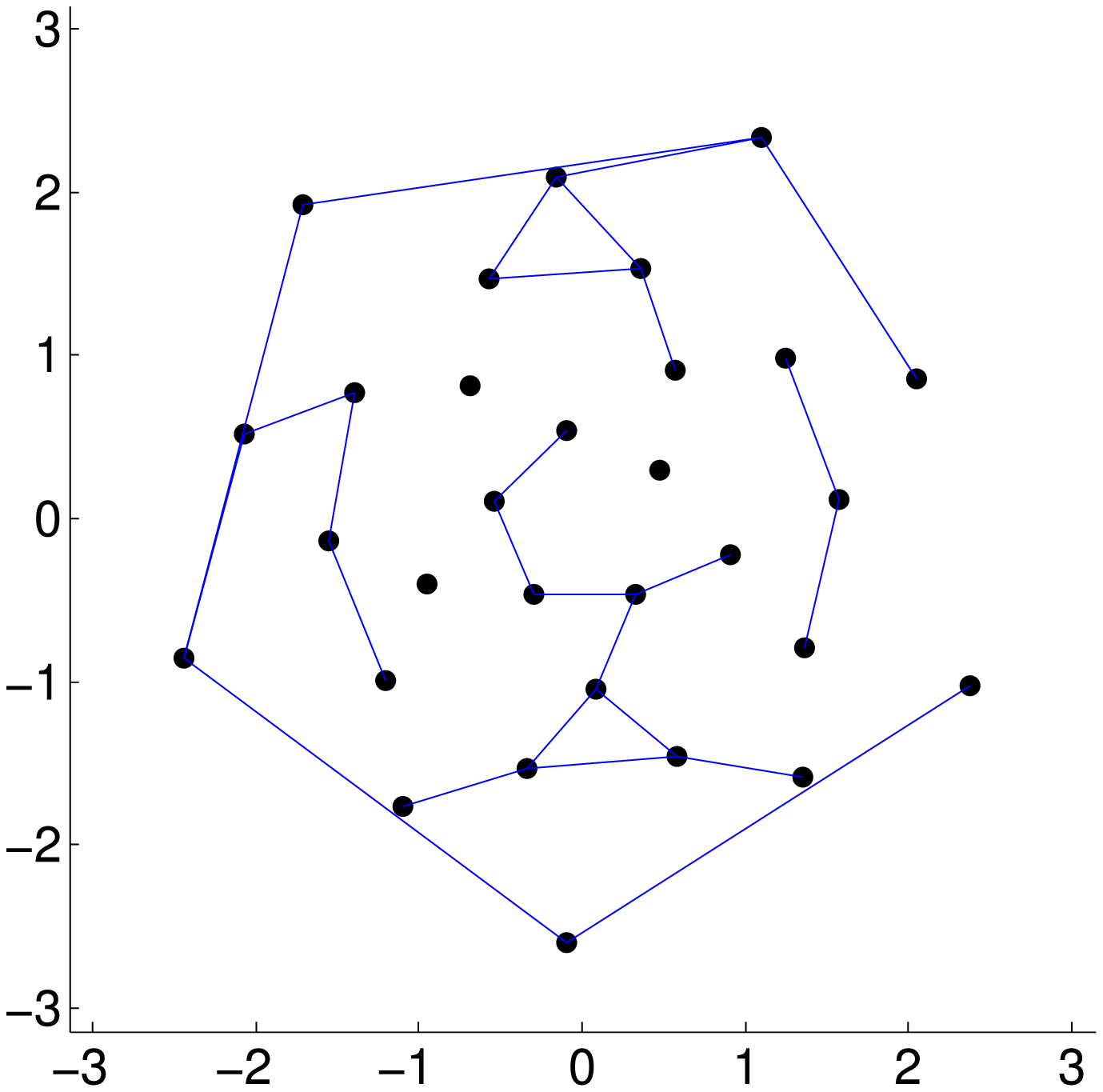}}
\caption{(Color online) The two shells of the $N=42$ DLP optimal packing; the
$12$ sphere centers of the first shell roughly form the vertices an icosahedron,
and the $30$ sphere centers (including $3$ rattlers) of the second shell roughly
form the vertices a icosidodecahedron. The zenith direction in both cases is
chosen along one of the rough $C_5$ axes. (a) $12$ spheres, $R = 1$, point group
$I_h$ (to $0.012$ sphere diameters). (b) $30$ spheres, $R_{min}(42) =
1.699423\dots$, point group $I_h$ (to $0.195$ sphere diameters).}
\label{N42}
\end{figure}

\begin{figure}[t]
\centering
\subfigure[]{\includegraphics[width=2.1in,clip]{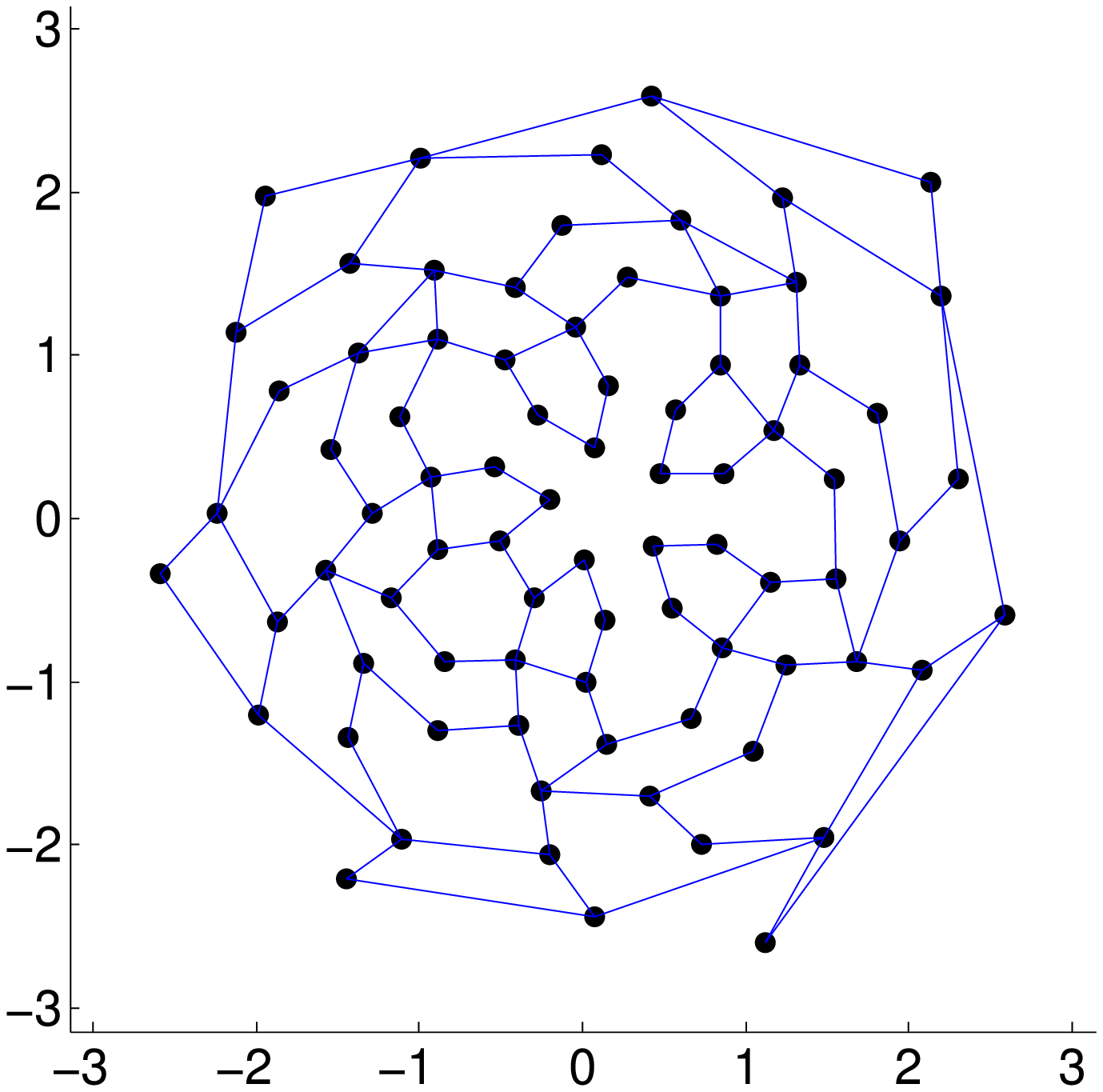}}
\caption{(Color online) The last shell of the roughly icosahedrally symmetric
$N=134$ DLP optimal packing, including $80$ spheres with centers at distance
$R_{min}(134) = 2.585816\dots$. The centers of $60$ of the spheres roughly form
the vertices of a truncated icosahedron, while the remaining $20$ spheres are
centered in the truncated icosahedron's hexagonal faces. The zenith direction is
chosen along one of the rough $C_5$ axes.}
\label{N134}
\end{figure}

The two shells of the $N=42$ DLP optimal packing are depicted in Fig. \ref{N42},
where the shells contain $12$ and $30$ spheres, respectively, just as do the
shells of the dense icosahedrally symmetric packing described in the first two
rows of Table \ref{icosTable}. The variation on the two perfectly icosahedrally
symmetric polyhedra that form the $N=42$ DLP optimal packing achieves an
improvement of $0.001879$ sphere diameters in $R$, with $R_{min}(42) =
1.699423\dots$. The last shell of the $N=134$ DLP optimal is depicted in Fig.
\ref{N134}. The $N=134$ packing includes; $12$ spheres within radial distances
$1$ and $1.036$, $30$ within radial distances $1.637$ and $1.763$, $12$ (of
which five are rattlers) within radial distances $1.999$ and $2.036$
(icosahedrally symmetric to $0.075$ sphere diameters), and $80$ at exactly
$R_{min}(134) = 2.585816\dots$. The DLP optimal packing for $N=134$ achieves an
improvement of $0.0197270\dots$ over the perfectly icosahedrally symmetric
packing.

\subsection{Maracas packings}
\label{maracasSec}
In a spherical region, the most area available to place the centers of
nonoverlapping spheres is on the surface. Consequently, is is perhaps expected
that DLP optimal packings would contain saturated or nearly saturated surfaces.
However, it is surprising that the salient features of certain DLP optimal
packings are entirely determined by the distribution of the spheres with centers
at precisely radius $R_{min}(N)$.

\begin{figure}[!ht]
\centering
\subfigure[]{\includegraphics[width=2.1in,clip]{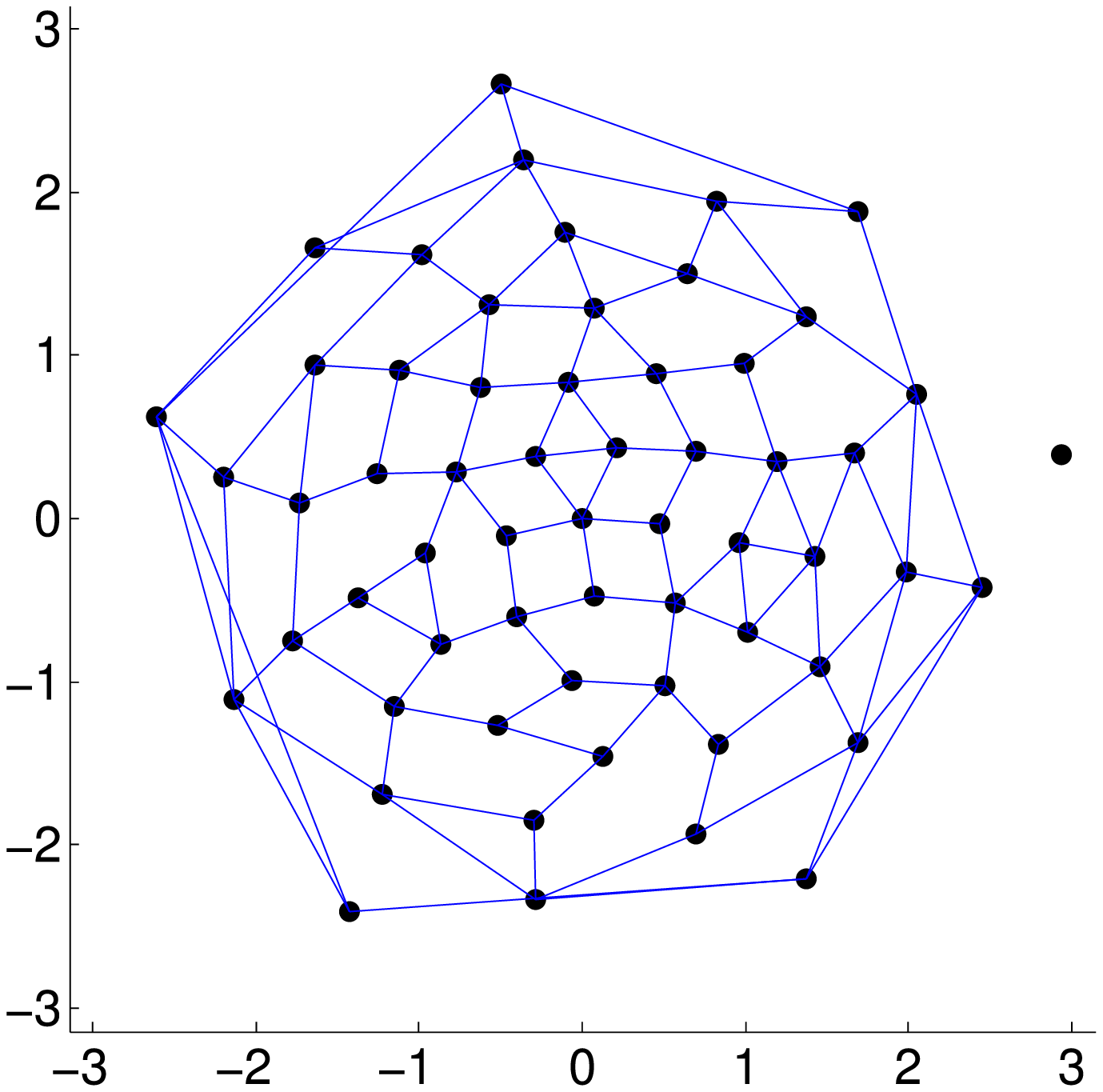}} \\
\subfigure[]{\includegraphics[width=2.1in,clip]{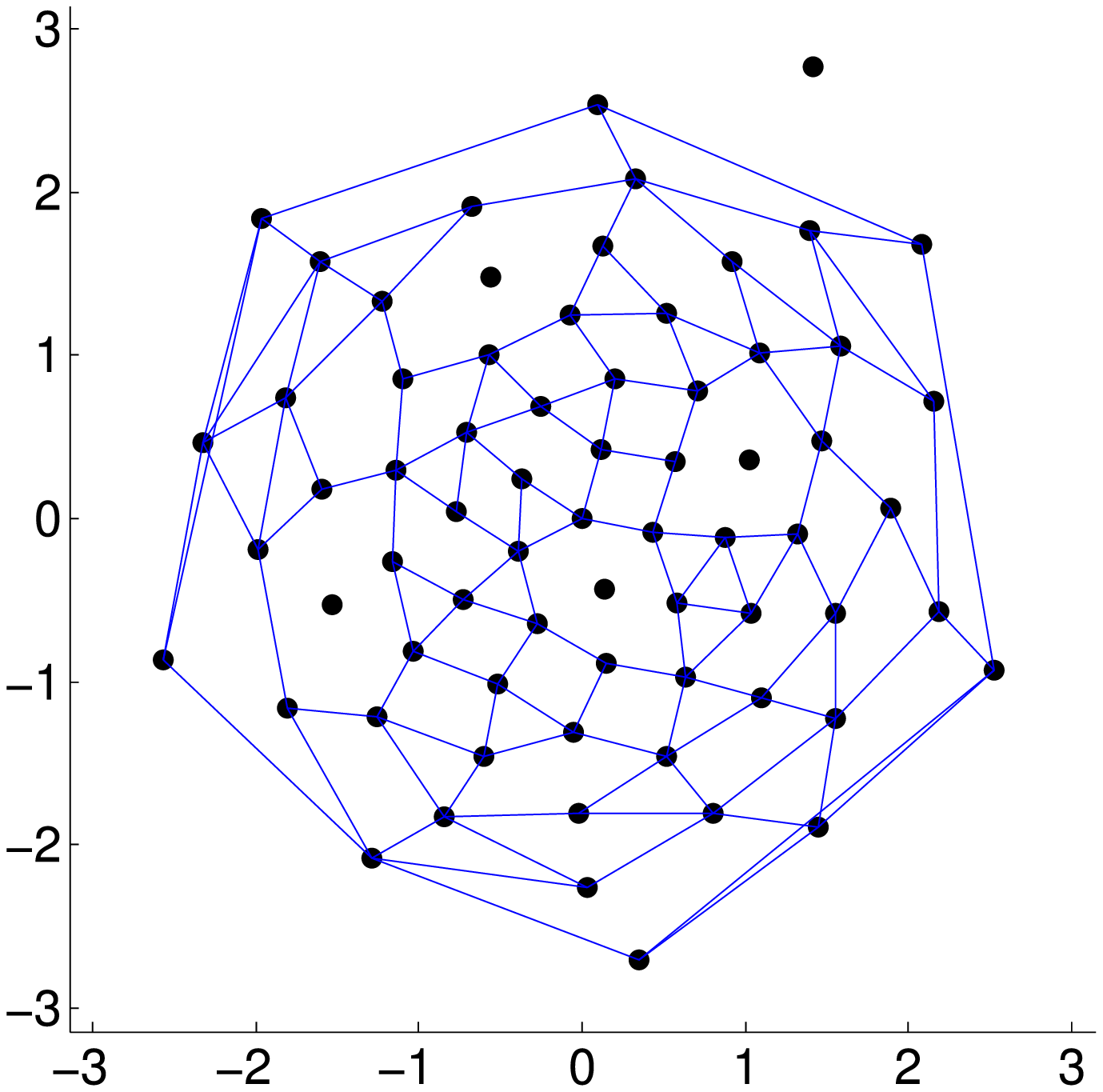}}
\caption{(Color online) The first and only shells for the $N=77$ and $N=93$
``maracas'' packings, including rattlers. The zenith direction in each case is
chosen somewhat arbitrarily to run through the center of one of the surrounding
spheres. (a) $59$ spheres, one rattler, $R_{min}(77) = 2.111526\dots$, point
group $C_1$. (b) $69$ spheres, five rattlers, $R = R_{min}(93) = 2.280243\dots$,
point group $C_1$.}
\label{maracas}
\end{figure}

The DLP optimal packings for $N=77$ and $N=93$ are termed ``maracas'' packings;
they are perfect examples of the phenomenon of surface-maximization, and exhibit
some of the lowest values of the greatest ${\mathcal S}$ compared, respectively,
to the packings in the sets ${\mathbb B}_{77}$ and ${\mathbb B}_{93}$. The
maracas packings each consist of a few unjammed spheres free to rattle within a
``husk'' composed of the maximal number of spheres that can be packed with
centers at $R_{min}(N)$. Further, $R_{min}(77) = R^S_{min}(N_{out}(77))$ and
$R_{min}(93)$ is only $6.606796\dots \times 10^{-5}$ sphere diameters larger
than $R^S_{min}(N_{out}(93))$. Figure \ref{maracas} depicts the first and only
shells for the maracas packings.

\subsection{Optimal packings where the central sphere is not locally jammed}
\label{cavity}
As in ${\mathbb R}^2$, for small enough $N$, there are many DLP optimal packings
in ${\mathbb R}^3$ where a number of jammed spheres form a cavity around the
central sphere such that were the central sphere not fixed, it would be a
rattler. The $N=62$ packing already discussed exhibits this characteristic,
containing eight spheres with centers at $R = 1.087542\dots$ and six with
centers at $R = 1.087786\dots$ from the center of the central sphere. The $N=61$
packing has a similar cavity composed of three layers of four spheres each with
centers at respective distances $R=1.013330\dots$, $R=1.028826\dots$, and
$R=1.019676\dots$, configured as squares. The $N=74$ optimal packing exhibits a
cavity composed of four layers of four spheres each with centers configured as
squares and at respective distances $R=1.152237\dots$, $R=1.156068\dots$,
$R=1.167618\dots$, and $R=1.225331\dots$. The cavities around the central sphere
in the $N=73$ and $N=78$ packings are composed of ten layers and nine layers,
respectively, of two spheres each.

The cavities formed are not, however, always symmetric. For $59$ of the $184$
$N$ studied, DLP optimal packings were found containing cavities such that the
center of the nearest sphere to the central sphere was at distance $R >1$; for
only seven of these is any layer composed of more than one sphere.

In general, the cavities range in number of spheres from $8$ for $N = 50$ to
$26$ for $N=99$. The center of the first sphere forming the cavity ranges from
distance $R=1.006188\dots$ for $N=154$ to $R = 1.356622\dots$ for $N=99$. The
farthest of the $26$ spheres forming the cavity in the $N=99$ DLP optimal
packing has center at distance $R=1.537500\dots$, indicating that the volume of
space available in the cavity to the center of the central sphere is more than
three times the sphere's volume.

\section{Conclusions}
\label{conclusions}
DLP optimal packings in ${\mathbb R}^3$ are widely spatially diverse and differ,
particularly on the surface, from subsets of the Barlow packings at all $N$.
They sometimes display elements of perfect symmetry and often display elements
of imperfect symmetry, such as imperfect icosahedral symmetry for sufficiently
small $N$. They are similar in these respects to DLP optimal packings in
${\mathbb R}^2$, which differ from packings of contacting disks with centers on
the vertices of the triangular lattice. However, at sufficiently large values of
$N$ in any ${\mathbb R}^d$, the bulk of DLP optimal packings must begin to
closely resemble a subset of one of the densest infinite packings in respective
${\mathbb R}^d$ or fail to be a densest local packing.

In ${\mathbb R}^3$, optimal packings tend to have a minimum number of shells and
a last shell that is almost always nearly saturated (or saturated). These
features lead to DLP optimal packings most closely resembling, as measured by a
scalar similarity metric (\ref{simMetric}), subsets of Barlow packings
consisting of $N+1$ spheres (a packing in ${\mathbb B}_N$) with the same
distribution of coordination shells as an FCC packing.

Knowledge of $R_{min}(N)$ for certain $N = N_*$ in ${\mathbb R}^d$ makes
possible the construction of a rigorous upper bound (\ref{upperBound}) on the
maximal density of an infinite sphere packing in ${\mathbb R}^d$; this bound
becomes more restrictive as $N_*$ grows large, and becomes the equality
$\hat{\phi}_*(N_*) = \phi_*^{\infty}$ as $N\rightarrow \infty$. Knowledge of the
$R_{min}(N)$ also makes possible the construction of a function $Z_{max}(R)$
that is an upper bound (\ref{ZRbound}) on the expected number of sphere centers
$Z(R)$ within distance $R$ from any given sphere center, which can be related to
a packing's pair correlation function $g_2(r)$ in ${\mathbb R}^3$ by (\ref{ZR}).
This upper bound is a realizability condition on a candidate pair correlation
function $g_2(r)$ for a packing of spheres, similar to the nonnegativity
conditions on $g_2(r)$ and its corresponding structure factor $S(k)$.

The function $Z_{max}(R)$ is also a significantly more restrictive upper bound
on candidate cross-correlation functions for a packing of a special central
sphere and its surrounding spheres, say, a single spherical solute molecule that
attracts same-size spherical solvent molecules. The critical distinction between
a cross-correlation function for a single sphere and a system-wide pair
correlation function $g_2(r)$ is that $g_2(r)$ is an expected value or average
over all identical spheres in a packing, whereas a cross-correlation function
for a single sphere applies locally, just as the function $Z_{max}(R)$ is
derived locally.

Considering, for example, the spheres forming the cavity wall in DLP optimal
packings in ${\mathbb R}^3$ with cavities around the central sphere, it is clear
that not every sphere in a packing can have the maximal number $Z_{max}(R)$ of
sphere centers within distance $R$ from its center (except for $R$ such that
$Z_{max}(R) = 12$, where the Barlow packings realize this criterion). However,
as any single (solute) sphere can have the maximal number of (solvent) sphere
centers within distance $R$ from its center, a $Z(R)$ defined in terms of a
cross-correlation function of a single central sphere can be equal to
$Z_{max}(R)$ at any $R$. Otherwise expressed, for $g_2^{yz}(r)$ any realizable
cross-correlation function between nonoverlapping spherical solute molecules of
unit diameter (of type $y$) in the dilute limit and same-size spherical solvent
molecules (of type $z$),
\begin{equation}
Z_{max}(R) = \sup \{\rho s_1(1)\int_0^Rx^{d-1}g_2^{yz}(x)dx\},
\label{crossZRbound}
\end{equation}
where the notation $\sup\{\,\dots\}$ indicates the mathematical supremum. The
function $Z_{max}(R)$ is thus a significantly more restrictive upper bound for
candidate cross-correlation functions of a single solute sphere amongst
same-size solvent spheres than for candidate pair correlation functions of
packings of indistinguishable spheres.

The characteristics of DLP optimal packings in each dimension $d$ are heavily
dependent on the underlying differences in packing spheres densely in ${\mathbb
R}^d$. For example, in ${\mathbb R}^2$ there is a unique arrangement of $K_2 =
6$ disks in contact with a central disk, whereas in ${\mathbb R}^3$ there are a
continuum of arrangements of $K_3 = 12$ spheres in contact with a central
sphere. In ${\mathbb R}^2$, there is a unique densest infinite packing of disks
and in ${\mathbb R}^3$, there are an uncountably infinite number of densest
infinite packings of spheres.

Differences in characteristics across dimension are also driven by the existence
of particularly locally dense dimensionally-unique packings. In ${\mathbb R}^2$,
these include the wedge hexagonal packings described in paper I and the curved
hexagonal packings \cite{LG1997a}, the densest local packings for $N = 3k(k+1)$,
$k=1,2,\dots6$. In ${\mathbb R}^3$, they include the two perfectly icosahedrally
symmetric packings for $N=42$ and $N=134$ (Table \ref{icosTable}) that contain
only two and four shells, respectively. It is curious to note that both the
curved hexagonal packings and the perfectly icosahedrally symmetric packings for
$N=42$ and $N=134$ are composed of a relatively small number of densely-packed
shells of spheres; however, in the curved hexagonal packings, the spheres in any
given shell are in contact with one another whereas in the aforementioned
perfectly icosahedrally symmetric packings, they are not.

Dimensions four, eight and twenty-four are similar to dimension two in that in
each of these dimensions there is a unique (uniqueness is conjectured for $d=4$
and proved for $d=8$ and $d=24$\cite{CSSPLG1998}) arrangement of spheres with
kissing numbers $K_4 = 24$ (recently proved by Musin in \cite{Musin2008a}),
$K_8=240$, and $K_{24}=196560$. The densest known packings in dimensions four,
eight, and twenty-four are also conjectured to be unique, and each is a lattice
packing that is self-dual, i.e., its reciprocal lattice is itself (the dual of
the triangular lattice is similarly a triangular lattice, though resized and
rotated $30$ degrees). The self-duality of the $E_8$ ($d=8$) and Leech ($d=24$)
lattices has been exploited to prove, up to a very small numerical tolerance,
that identical nonoverlapping spheres with centers on the sites of these
lattices are the densest packings of spheres in their respective dimensions
\cite{CE2003a}. In dimension five, the densest known packings can be described,
similarly to Barlow packings, as stackings of layers of the densest packings in
${\mathbb R}^4$. Consequently, we might expect to find DLP optimal packings in
${\mathbb R}^d$, $d=4$, $8$, and $24$ similar to the curved hexagonal packings
in ${\mathbb R}^2$, whereas we would not expect to find such optimal packings in
${\mathbb R}^5$.

These and other dimension-dependent dense packing characteristics could have an
effect on the probability of freezing in a overcompressed liquid of hard spheres
in ${\mathbb R}^d$. Recent work \cite{SBM2006a,SBM2007a,SSOOS2010a,KT2010a}
suggests that the phase transition from a overcompressed hard-sphere liquid in
${\mathbb R}^3$ to a crystalline solid with sphere centers near to the sites of
the spheres centers in a Barlow packing may be described as a two-stage process.
In the first stage, small clusters of spheres form that are denser than either
the liquid or crystalline solid states. In the second stage, the dense clusters
grow in size and decrease in density while their bulk (interior) transforms from
the center outward into a crystalline solid state \cite{SSOOS2010a}. An analogy
can be drawn between DLP optimal packings in ${\mathbb R}^3$ for smaller $N$ and
the small clusters, in that both are locally denser than corresponding subsets
of Barlow packings or crystalline solid states and in that neither are similar
(both angularly and radially) to small subsets of Barlow packings. A similar
analogy applies between DLP optimal packings for sufficiently large $N$ and the
larger clusters with crystalline solid interiors, in that both are very similar
to Barlow packings in their bulk and not similar on and near their surfaces.

In general, the probability per unit time of freezing in hard-sphere liquids at
comparable overcompression decreases with increasing dimension, at least for
dimensions two through six \cite{MCFC2009a}. However, with the previously
described two-stage process in mind, consideration of the dimension-dependent
characteristics of the densest local packings could lead to a further increase,
beyond what might otherwise be predicted by the general trend, in estimates of
the probability per unit time of freezing in ${\mathbb R}^2$ as compared to in
${\mathbb R}^3$.

For example, unlike in ${\mathbb R}^3$, in ${\mathbb R}^2$ there are several
values of $N > K_2 = 6$, specifically, $N=12$, $30$ and $54$, for which one of
the densest local packings is a subset of the densest infinite packing, and
therefore for which the equality in the upper bound (\ref{ZRbound}) can hold. We
might expect to see this occurring in ${\mathbb R}^4$, ${\mathbb R}^8$, or
${\mathbb R}^{24}$ as well, but not in ${\mathbb R}^5$. The equivalence of
densest local packings at certain $N$ and subsets of the densest infinite
packing in ${\mathbb R}^2$ suggests that the first stage of the two-stage
crystallization process, in which small clusters of spheres form that are denser
than either the liquid or crystalline solid states, may be shortened in duration
for hard-disk liquids in ${\mathbb R}^2$ relative to hard-sphere liquids in
${\mathbb R}^3$. If this is the case, accounting for the equivalence of densest
local packings and subsets of the densest infinite packings should result in an
increase in the probability per unit time of freezing in ${\mathbb R}^2$ and
potentially ${\mathbb R}^4$, ${\mathbb R}^8$, and ${\mathbb R}^{24}$ as compared
to in ${\mathbb R}^3$ and ${\mathbb R}^5$. For hard-sphere liquids at densities
near the freezing point, in ${\mathbb R}^2$ relative to in ${\mathbb R}^3$, the
more pronounced ``shoulder'' \cite{TTSDS1998a} appearing in pair correlation
functions between the first and second nearest-neighbor distances could be
evidence indicating such a shortened first stage.

This example and the similarities between DLP optimal packings and the nuclei
described in the two-stage description of crystallization suggest that there may
be an explicit connection between freezing in overcompressed hard-sphere liquids
and DLP optimal packings. It would be interesting, in the context of a revised
nucleation theory, to explore the relationship between nucleation and dense
local clusters of spheres configured with centers near the sites of sphere
centers in the densest local packings.

\bigskip
\noindent {\bf ACKNOWLEDGEMENTS:}
\medskip

S.T. thanks the Institute for Advanced Study for its hospitality during his stay
there. This work was supported by the Division of Mathematical Sciences at the
National Science Foundation under Award Number DMS-0804431 and by the MRSEC
Program of the National Science Foundation under Award Number DMR-0820341.

\appendix
\section{Lower bounds on $R_{min}(N)$}
\label{lowerBound}

Knowledge of the densest infinite packings in ${\mathbb R}^3$, the Barlow
packings, allows $R_{min}(N)$ to be bounded both from above and below. A lower
bound can be obtained through the observation that it is not possible to remove
a finite number $N+1$ of spheres from an infinite Barlow packing and replace
them by $N+1$ spheres packed as an optimal DLP packing. If this could be
accomplished at even one value of $N$, then the Barlow packings would not be the
only densest infinite packings of identical nonoverlapping spheres in ${\mathbb
R}^3$, as this operation could be repeated {\it ad infinitum} to yield an
infinite-volume packing fraction $\phi_*^{\infty}$ greater than or equal to
$\pi/\sqrt{18}$.

Consider removing all $N+1$ spheres (including a central sphere) in a Barlow
packing with centers within distance $R(N) = R_{min}(N) + \epsilon$ of the
center of the central sphere, where $N = N'$ is chosen to include all spheres in
the coordination shell at $R(N')$. We term the set of all such subsets of $N'+1$
spheres chosen from all Barlow packings ${\mathbb B}_{N'}$, similar to the
${\mathbb B}_N$ defined in Sec. \ref{SCandBarlow}. Attempting to replace
(without overlap) the removed spheres with a DLP packing of $N'+1$ spheres with
optimal radius $R_{min}(N')$, we see that $\epsilon$ must be in the range $0
\leq \epsilon \leq 1$. If $\epsilon$ were less than $0$, then the DLP packing
wouldn't be optimal. If $\epsilon$ were greater than or equal to $1$, then the
Barlow packings would not be the only densest packings of identical
nonoverlapping spheres in ${\mathbb R}^3$, as for $\epsilon \geq 1$, the DLP
packing with optimal radius $R_{min}(N')$ can be always be placed such that none
of its spheres overlaps any of the remaining spheres in the Barlow packing. This
range of epsilon results in the lower bound $R_{min}(N') \geq R(N') - 1$, valid
for any $R(N') \geq R_{min}(N')$ chosen as stated above.

In practice, $\epsilon$ may be reduced significantly below the value of $1$. For
example, as for certain $N'$, $R(N') = R_{min}(N') + \epsilon$ varies between
the packings in the set ${\mathbb B}_{N'}$, $\epsilon$ is reduced from unity by
the difference between the largest and smallest $R(N')$ for these $N'$. More
generally, the value of $\epsilon$ can be significantly reduced by investigating
the geometric considerations of placing the $N+1$ spheres of any DLP optimal
packing of radius $R_{min}(N)$ into the void created by removing $N$ spheres
from a Barlow packing. Reducing $\epsilon$ to a minimum possible value
$\epsilon_{min}(N)$ results in the lower bound $R_{min}(N) \geq R(N) -
\epsilon_{min}(N)$.

\section{Barlow packings and similarity metric reference sets}
\label{finiteBarlowPackings}

Any set of $N+1$ spheres in a given ${\mathbb B}_N$ can be used as a reference
set in the similarity metric (\ref{simMetric}). However, for $N>0$, there are
always multiple reference sets of $N+1$ spheres that will produce the same
$\{\delta_i\}$ and consequently the same value of ${\mathcal S}$ for a given
comparison set. For example, as the metric (\ref{simMetric}) is insensitive to
the angular position of any sphere, for a furthest coordination shell that is
not full, the same $\{\delta_i\}$ is produced regardless of which of the spheres
in the furthest shell is included in the packing used as a reference set.

When these and all other degeneracies are taken into account, we find that for
${\mathbb B}_N$ including only packings of $s=1,3,5,7,9,11$, and $13$ Barlow
stacking layers, where all ${\mathbb B}_{1054}$ packings are constructed from
$13$ layers, there are $1,1,3,6,14,31$ and $70$ distinct sets $\{\delta_i\}$.
For the $N$ studied where the packings in ${\mathbb B}_N$ do not all have the
same number of layers, the packings in ${\mathbb B}_N$ include $s$, $s+1$, and
sometimes $s+2$ layers. For these ${\mathbb B}_N$, the set of distinct
$\{\delta_i\}$ is a subset of the set of distinct $\{\delta_i\}$ for any
${\mathbb B}_N$ including packings of only $s+2$ layers, and the number of
distinct sets $\{\delta_i\}$ is between the number for ${\mathbb B}_N$ including
only packings of $s$ layers and for ${\mathbb B}_N$ including only packings of
$s+2$ layers.

An intuitive analysis of the determination of the shells $\{\delta_i\}$ suggests
that the presence of relatively fewer shells in a reference set can generally
increase the value of ${\mathcal S}$. An increase of this sort does occur when
shells are of radial width large enough such that the number of sphere centers
within each shell approaches the number density of the packing, i.e., when
radial width is on the order of a sphere's diameter. However, this is not the
case for the $\{\delta_i\}$ derived from the sets in each ${\mathbb B}_N$.
Indeed, direct analysis of the distribution of spheres within each individual
shell $\delta_i$ confirms that spheres in DLP optimal packings for a given $N$
are in general radially configured in a relatively small number of shells. That
is, the radial positions of the sphere centers in DLP optimal packings are
clustered around a smaller number of distances from the center of the central
sphere, relative to the number of coordination shells in the packings in
${\mathbb B}_N$, and the large fraction of DLP optimal packings that are most
similar to FCC-derived packings is not the result of a design-flaw in the
similarity metric (\ref{simMetric}). 


\begin{thebibliography}{42}
\bibitem[{end({\natexlab{a}})}]{endnote4}
\bibinfo{note}{In practice, we find over the range of $N$ studied in ${\mathbb
  R}^3$ that the only $N > 12$ for which $R_{min}(N) = R_{min}(N-1)$ is
  $N=60$.}
\bibitem[{end({\natexlab{b}})}]{endnote2}
\bibinfo{note}{For $N=216$, $Z^S_{max}(R_{min}(216)) = 130$, and $130$ is the
  last number for which optimal spherical codes are presented in
  \cite{SloaneSCweb}.}
\bibitem[{end({\natexlab{c}})}]{endnote5}
\bibinfo{note}{The Barlow packings consist of translated planes (layers) of
  contacting spheres with centers arranged on the sites of a triangular
  lattice, where the layers are ``stacked'' on top of one another. In the
  stackings, there are three possible translations relative to the central
  layer (the layer including the central sphere), denoted $A$, $B$, and $C$,
  obeying the rule that no layer can be consecutively repeated. We have found
  that it is only the positions of the $A$ layers relative to the central layer
  that are relevant in determining the coordination shells about the central
  sphere. The $FCC$ Barlow stacking, denoted by repeating $ABC$ layers, is thus
  indistinguishable to the similarity metric from any other stacking with an
  $A$ layer repeated every three layers.}
\bibitem[{end({\natexlab{d}})}]{endnote1}
\bibinfo{note}{For $N=42$, the $V_{LJ}$ of the DLP and Lennard-Jones optimal
  configurations are $-189.63125$ and $-202.36466$ \cite{CambridgeCluster},
  respectively, and for $N=114$, the $V_{LJ}$ are $-634.02115$ and $-655.75631$
  \cite{CambridgeCluster}, respectively.}
\bibitem[{end({\natexlab{e}})}]{endnote3}
\bibinfo{note}{In the diagrams of paper I, e.g., for DLP optimal packings in
  ${\mathbb R}^2$ for $N=24$, $N=45$ and $N=95$ disks, it is difficult to
  perceive that the disks forming a cavity around the central disk are not all
  the same radial distance from the central disk. However, though the
  difference in radial distances between these disks is small, it is far larger
  than the precision to which the spatial positions of sphere centers are
  calculated by the algorithm.}
\end{thebibliography}

\end{document}